\documentclass[manuscript,screen]{acmart}

\usepackage{listings}
\usepackage{tabularx}
\usepackage{hyperref}
\usepackage{epigraph} 
\usepackage{dirtytalk}

\AtBeginDocument{%
  \providecommand\BibTeX{{%
    \normalfont B\kern-0.5em{\scshape i\kern-0.25em b}\kern-0.8em\TeX}}}

\setcopyright{acmcopyright}
\copyrightyear{2022}
\acmYear{2022}
\acmDOI{XXXXXXX.XXXXXXX}

\acmJournal{JACM}
\acmMonth{12}

\begin{document}

\title{Quantifying Technical Debt: A Systematic Mapping Study and a Conceptual Model}

\author{Judith Perera}
\email{jper@aucklanduni.ac.nz}
\authornotemark[1]
\affiliation{
  \institution{The University of Auckland}
  \city{Auckland CBD, Auckland}
  \country{New Zealand}
  \postcode{1010}
}

\author{Ewan Tempero}
\email{e.tempero@auckland.ac.nz}
\affiliation{
  \institution{The University of Auckland}
  \city{Auckland CBD, Auckland}
  \country{New Zealand}
    \postcode{1010}
}

\author{Yu-Cheng Tu}
\email{yu-cheng.tu@auckland.ac.nz}
\affiliation{
  \institution{The University of Auckland}
  \city{Auckland CBD, Auckland}
  \country{New Zealand}
    \postcode{1010}
}

\author{Kelly Blincoe}
\email{k.blincoe@auckland.ac.nz}
\affiliation{
  \institution{The University of Auckland}
  \city{Auckland CBD, Auckland}
  \country{New Zealand}
    \postcode{1010}
    }

\begin{abstract}
To effectively manage Technical Debt (TD), we need reliable means to quantify it. We conducted a Systematic Mapping Study (SMS) where we identified TD quantification approaches that focus on different aspects of TD. Some approaches base the quantification on the identification of smells, some quantify the Return on Investment (ROI) of refactoring, some compare an ideal state with the current state of a software in terms of the software quality, and some compare alternative development paths to reduce TD. It is unclear if these approaches are quantifying the same thing and if they support similar or different decisions regarding TD Management (TDM). This creates the problem of not being able to effectively compare and evaluate approaches. 
To solve this problem, we developed a novel conceptual model, the Technical Debt Quantification Model (TDQM), that captures the important concepts related to TD quantification and illustrates the relationships between them. TDQM can represent varied TD quantification approaches via a common uniform representation, the TDQM Approach Comparison Matrix, that allows performing useful comparisons and evaluations between approaches. 
This paper reports on the mapping study, the development of TDQM, and on applying TDQM to compare and evaluate TD quantification approaches.
\end{abstract}



\maketitle

\section{Introduction}

Technical Debt (TD) captures the consequences of making sub-optimal decisions during software product development \cite{Cunningham1992}. Taking TD can be beneficial in the short-term, for example, to deliver a product in time to market. However, in the long run, it could become detrimental to the software product as the quality of the product degrades due to the sub-optimal decisions made during development \cite{behutiye2017analyzing}. Then it becomes difficult to add new features to the product having to do extra work to be able to implement new features while there is TD \cite{behutiye2017analyzing, besker2017pricey}. Therefore, TD must be managed during software development \cite{tom2013exploration, fowler_2019}. However, effective TD Management (TDM) relies on the ability to quantify it ---  \emph{\say{In order to manage technical debt, a way to quantify the concept is needed}} \cite{GuoandSeaman2011}. We are investigating how TD can be quantified. In this paper, we look at what proposals have been made to quantify TD and how TD quantification can be modelled to better evaluate those proposals. 

We conducted a \emph{Systematic Mapping Study (SMS)} to answer the research question, \emph{\textbf{"RQ1: What approaches to TD quantification have been proposed in the research literature?"}}. The mapping study results showed that various approaches to quantifying TD proposed in the literature focus on different aspects of TD. For example, some approaches base their quantification on the identification of smells, e.g., identification of code, design, or architecture smells \cite{Sharma2016, Fontana2017, zazworka2011prioritizing, verdecchia2020atdx, roveda2018towards, charalampidou2017assessing, fontana2015towards}, some approaches quantify the Return on Investment (ROI) of refactoring \cite{Kazman2015}, some compare an ideal state with the current state of a software in terms of the quality of the software \cite{Singh2014, Reimanis2016, Letouzey2016, von2019mitigating, nugroho2011empirical, martini2018semi} and some compare alternative development paths to reduce the accumulation of TD during software development \cite{Nord2012, Sangwan2020}. It is unclear if such approaches are quantifying the same thing and if they support similar or different decisions regarding TDM; leading to the problem of not being able to compare and evaluate quantification approaches effectively as well as efficiently.

To address this problem, we developed the \emph{Technical Debt Quantification Model (TDQM)} --- a novel conceptual model that models the quantification of TD by capturing the important concepts related to TD quantification and illustrating the relationships between these concepts. The development of the model answered the research question, \emph{\textbf{"RQ2: How can we model TD quantification?"}}. TDQM can represent the varied TD quantification approaches via a common uniform representation, the \emph{TDQM Approach Comparison Matrix}, that allows performing useful comparisons and evaluations between quantification approaches. 

We developed TDQM to model the quantification of types of TD related to software code such as \emph{Code, Design and Architectural TD}. It was developed based on the knowledge obtained via examining existing literature. The model was then used to compare and evaluate quantification approaches for the same code-related types of TD found in our mapping study before and after the development of the model, answering the research question,\emph{\textbf{"RQ3: How can we compare and evaluate quantification approaches?"}}. In our future work we will investigate the possibility of extending TDQM to other types of TD such as \emph{Requirements TD}.

This paper reports on the methodology and results of the mapping study, presents TDQM and its development and then reports on the methodology and results of the study conducted to compare and evaluate quantification approaches. The main contributions of this paper can be summarized as follows;

\begin{itemize}
    \item A Systematic Mapping Study (SMS) --- answering \textbf{RQ1}
    \item A Conceptual Model --- \emph{Technical Debt Quantification Model (TDQM)} that captures important concepts related to TD quantification and illustrates relationships between them, and the \emph{TDQM Comparison Matrix} which is a common uniform representation that enables effective and efficient comparisons and evaluations of quantification approaches --- answering \textbf{RQ2}
    \item A study applying TDQM to compare and evaluate quantification approaches found in our mapping study before and after the development of the model --- answering \textbf{RQ3}
\end{itemize}

The paper begins by providing background and discussing related work in Section \ref{sec:related_work}, then reporting on the Systematic Mapping Study in Section \ref{sec:sms_method} (SMS Methodology) and Section \ref{sec:sms_results} (SMS Results). In Section \ref{sec:method:tdqm} (TDQM Methodology), the paper reports on the development of TDQM and presents the model in Section \ref{sec:results:tdqm} (TDQM Results) as well as illustrates the application of TDQM using two hand-picked examples from the literature. Afterward, in Section \ref{sec:method:tdqm_app} (Comparing and Evaluating Methodology) the paper reports the methodology  and in Section \ref{sec:results:tdqm_app} (Comparing and Evaluating Results) the results of applying TDQM to quantification approaches. Section \ref{sec:discussion} discusses our findings and the contributions and limitations of TDQM as well as threats to validity, future work and implications for researchers and practitioners. The paper is concluded in Section \ref{sec:conclusion}.

\section{Related Work}
\label{sec:related_work}

\subsection{Secondary and Tertiary Studies}

Several secondary and tertiary studies related to TD have been published starting from 2012. However, to the best of our knowledge, our mapping study is the first to bring \emph{`TD Quantification'} to the limelight as opposed to the published secondary and tertiary studies described below. 

\subsubsection{Concept of TD}

Tom et al. [6] were the first to conduct a secondary study on TD in 2012. They focused on the concept of TD, attempting to provide a holistic view of TD. The outcome was a theoretical framework comprising TD dimensions, attributes, precedents, and outcomes. Alves et al. [7] proposed an ontology of terms on TD in their systematic literature review. Different types of TD and their indicators were identified during this study. Alves et al. [11] provided an improved version of the
ontology of terms on TD proposed by the same authors in 2014. They provided a list of indicators to identify TD, a list of TD management strategies, data sources used in TD identification activities, and software visualization techniques used to identify and manage TD. 

\subsubsection{TD Management (TDM), TDM Tools and Strategies}

Li et al. [8] conducted their study on TD and TDM. They classified TD into ten types and TDM activities into eight types. Twenty-nine tools for TDM were identified. The authors emphasized the need for tools for managing the different TD types during the TDM process. Rios et al. [5] conducted a tertiary study in 2018. They evaluated 13 secondary studies between 2012 and 2018. They consolidated the TD types found in previous secondary studies, identified a list of situations in which TD items can be found in software projects, and presented a map representing the activities, strategies, and tools supporting TDM. They pointed out TDM activities that do not yet have any support tool. Khomyakov et al. [15] investigated existing tools for the measurement of TD, but they focused only on quantitative
methods that could be automated. They reported on 38 papers out of 835 retrieved in their initial search. Avgeriou et al. [16] compared a few existing tools measuring TD. They compared the features and the popularity of the tools. They focused on the TD types: code, design, and architecture. Behutiye et al. [13] analyzed the state of the art of TD and its causes, consequences, and management strategies in the context of agile software development (ASD).

\subsubsection{Decision Making in TD Management}

Fernández-Sánchez et al. [12] identified elements required to manage TD. The elements were classified into three groups: basic decision-making factors, cost estimation techniques, practices, and techniques for decision-making. The factors were grouped based on stakeholders’ points of view: engineering, engineering management, and business-organizational management. Ribeiro et al. [10] evaluated the appropriate time for paying a TD item and how to apply decision-making criteria to balance the short-term benefits against long-term costs. They identified 14 decision-making criteria that development teams can use to prioritize the payment of TD items and a list of types of debt related to the criteria. 

\subsubsection{TD Prioritization}

Alfayez et al. [17] investigated TD prioritization approaches and the prioritization techniques utilized by those approaches. Furthermore, they analyzed prioritization approaches based on their accounts for value, cost, or resource constraints. Leanarduzzi et al. reviewed articles on technical debt prioritization including strategies, processes, factors, and tools. They discovered that there is a lack of empirical evidence on measuring TD and that there is no validated, widely used set of tools specific to TD prioritization.

\subsubsection{Financial aspect of TD}

Ampatzoglou et al. [9] focused on the financial aspect of TD. The authors provided a glossary of financial terms and a classification scheme for financial approaches to managing TD. 

\subsubsection{Architectural TD (ATD)}

Besker et al. [14] investigated Architectural TD (ATD) in their systematic literature review. They provided a comprehensive interpretation of the ATD phenomenon by contributing with a descriptive model categorizing the main characteristics of ATD.

The existing secondary and tertiary studies have not emphasized enough on \emph{`TD Quantification'} or measurement as a concept. In our mapping study, we emphasize the importance of understanding TD quantification as a concept as well as an important TDM activity since it supports other TDM activities such as \emph{`prioritization'} and \emph{`repayment'} as well \cite{MARTINI2015237, 8786030}. 

\subsection{Technical Debt Management and Quantification of Technical Debt}

The first mention of TD as a metaphor to indicate writing not quite-right code as a trade off of long-term code quality for a short-term gain, was by Cunningham in 1992 \cite{Cunningham1992}. Avgeriou et al. proposed a consensus definition for TD at a Dagstuhl seminar held in 2016, referred to as the 16162 definition of TD: \emph{\say{In software-intensive systems, technical debt is a collection of design or implementation constructs that are expedient in the short term, but set up a technical context that can make future changes more costly or impossible. Technical debt presents an actual or contingent liability whose impact is limited to internal system qualities, primarily maintainability and evolvability \cite{avgeriou2016managing}.}}

Avgeriou et al. \cite{avgeriou2016managing} also introduced a conceptual model for TD (referred to as the '16162 model' in this paper) based on two viewpoints. The \emph{first viewpoint} describes the properties, artifacts, and elements related to technical debt items, and the \emph{second viewpoint} articulates the management and process-related activities or the different states that debt may go through. In the 16162 model, TD was described as one of many concerns in a software system. The authors also discussed the concept of a TD item. A TD item is associated with one or more artifacts of the software development process, such as code, test, or documentation, and is caused by, for example, schedule pressure. Our model utilizes the concept of a TD Item.

The 16162 model does not entirely capture the two viewpoints. Instead, it focuses more on the first viewpoint, capturing the elements related to TD. Therefore, the model does not discuss quantification (or measurement) of TD, which is one of the Technical Debt Management (TDM) activities introduced by Li et al. \cite{Li2015} in a previous study. According to Li et al., TDM includes activities that prevent potential TD from being incurred (e.g., prevention) and activities that deal with accumulated TD to make it visible, controllable and to keep a balance between costs and value of a software project (e.g., identification, visualizing, monitoring, measurement, prioritization, repayment).

Li et al. \cite{Li2015} and Rios et al. \cite{Rios2018a} (a more recent study) identify \emph{'measurement'} as a key TDM activity. In our paper, we describe the same TDM activity as \emph{'quantification'}. We focus on 'quantification' rather than the rest of the TDM activities since TDM is hindered by the inability to quantify TD usefully \cite{GuoandSeaman2011}. Additionally, TDM activities such as \emph{'prioritization'} and \emph{'repayment'} too rely on the \emph{quantification} of TD \cite{MARTINI2015237, 8786030}. Avgeriou et al. \cite{avgeriou2020overview} provide an overview of the current state of the market for TD measurement tools. However, their study is limited to tools that estimate TD principal or interest. Our research complements their research by extending the discussion of TD quantification to a greater degree without limiting it to principal and interest.

Ribeiro et al. \cite{ribeiro2016decision} introduce criteria that can be utilized for TDM decision making. 
They identify \emph{'Debt impact on the project'} and \emph{'Cost-Benefit'} as the most explored criteria in the studies captured in their mapping study. This confirms our selection of concepts (i.e., concepts related to Cost and Benefit, including TD Interest) to build our theory for discussing TD Quantification.

We emphasize on the need to understand the important concepts related to TD quantification and their relationships to be able to better comprehend existing proposals made for TD quantification. We identified the problem of \emph{not being able to compare and evaluate existing proposals quantifying TD} as there was \emph{no consensus among these approaches as to what they were quantifying and what TDM decisions they were supporting}. As a result of the identified problem, we developed a conceptual model that serves as a reference point to better comprehend, compare, and evaluate quantification approaches. Our model captures the important concepts related to TD quantification and illustrates the relationships between them. We then utilize this model to compare and evaluate quantification approaches found in our mapping study.

\section{RQ1: Exploring approaches to TD quantification --- a Systematic Mapping Study --- Methodology}
\label{sec:sms_method}

We followed recommendations given by Kitchenham et al. \cite{kitchenham2010systematic} and Petersen et al. \cite{petersen2008systematic} to conduct a Systematic Mapping Study, which is a form of Systematic Literature Review (SLR) --- \emph{\say{a methodologically rigorous review of research results}} as described by Kitchenham et al. The research question for the Systematic Mapping Study (SMS) was; 

\begin{itemize}
    \item \emph{\textbf{RQ1: What approaches to TD quantification have been proposed in the research literature?}}
\end{itemize}

Papers were gathered in two iterations; \emph{Iterations 1 and 2}, where 113 and 16 primary studies were retrieved, respectively. Following a rigorous searching and screening process, we queried five digital databases \emph{(SCOPUS, IEEE, ACM, SpringerLink, ScienceDirect)} in both iterations. The second iteration of the mapping study was updating the mapping study data set. The primary studies found in the second iteration also served as a validation set of data, for the validation of our model developed to answer RQ2 in Section \ref{sec:results:tdqm}. The model was developed before the second iteration of the mapping study and did not require any changes when the mapping study was updated. Below, we briefly describe the methodology for our mapping study. 

\subsection{Search Strategy}

Articles from an initial manual search using studies from ICSE TechDebt conferences 2018 and 2019 were used as a reference set to build our search string. Search terms were extracted from the title, abstract, keywords, and sections of the full text where they seemed appropriate. After that, the search terms were expanded with synonyms. We tested the search phrase in SCOPUS, which covered all the articles in our reference set. Hence, we could obtain 100 percent precision of our preliminary search string. The search string contained the following terms and synonyms in its final version:\emph{ Technical Debt, quantify, measure, forecast, predict, assess, estimate, calculate, amount,
value, impact, principal, interest, metric, time, cost}. We used the term ‘Technical Debt’ along with the rest of the keywords since it helps us scope down and avoid articles that do not focus on technical debt but software quality or architecture alone. Digital databases used for obtaining primary studies, IEEEXplore, ACM, and Science Direct, were recommended by Brereton et al. \cite{brereton2007lessons}, while SCOPUS was recommended by Cavacini \cite{cavacini2015best}. We used the asterisk character \emph{(*)} to capture possible keyword variations, for example, plurals and verb conjugations. We applied the query to the title, abstract, and keywords to increase the probability of finding all relevant publications. The final Search String for SCOPUS is shown in Listing \ref{listing_search_str}. We tailored this search string for the rest of the digital databases according to the functionality and usability of their interfaces.

\begin{center}
\begin{lstlisting}
TITLE-ABS-KEY ( "Technical Debt" 
	AND ( quantif* 
		OR measur* 
		OR forecast* 
		OR predict* 
		OR assess* 
		OR estimat* 
		OR calculat* 
		OR impact* 
		OR amount 
		OR valu* 
		OR principal 
		OR interest* 
		OR metric* 
		OR time 
		OR cost* ) )
	AND ( LIMIT-TO ( DOCTYPE , "cp" ) 
		OR LIMIT-TO ( DOCTYPE , "ar" ) ) 
	AND ( LIMIT-TO ( LANGUAGE , "English" ) )
\end{lstlisting}
\captionof{lstlisting}{Final Search String for SCOPUS}
\label{listing_search_str}
\end{center}

\subsection{Article Screening and Selection}

Article screening and selection was performed by applying the inclusion/exclusion criteria listed in Table \ref{tab:inclexcl}. We followed the \emph{adaptive depth reading approach} for screening articles as suggested in Petersen et al. \cite{petersen2008systematic} starting from the title and then continuing through the abstract, conclusion, and at last, reading the full text. 

Articles that described an approach to quantifying TD, either introducing or evaluating an approach, were included. Articles that described quantifiable characteristics of TD (e.g., principal, interest, interest probability) or units of measurement (in terms of; time, cost, or effort) were included. Since we were interested in the applicability of software metrics in the measurement of TD, we included articles discussing software metrics concerning TDM. 

We did not consider secondary or tertiary studies as they would count the primary studies multiple times. We included only peer-reviewed articles as they are considered of high quality. We ruled out articles not written in English since the authors were not confident in other languages. Research articles not directly related to quantifying TD, i.e., articles describing other TDM activities and not quantifying TD, and articles related to software quality measurement but not related to TD, were ruled out. Articles that we could not access the full text were too ruled out.

The first author screened the articles. When doubt was encountered, they were recorded and then discussed and resolved during discussions with the other authors. The inclusion and exclusion criteria were well tested and agreed upon after refining them in a few iterations during the initial manual search phase conducted on the reference set of papers obtained from TD conferences 2018 and 2019.

\begin{table}[h]
\centering
\begin{tabularx}{\textwidth}{c | l} 
	Inclusion Criteria & \\
	  	& I1 Discusses approaches quantifying TD\\
	  	& I2 Discusses quantifiable characteristics of TD\\
	  	& I3 Discusses units of measurement that could be used to quantify TD\\
	  	& I4 Discusses SW Metrics in relation to TDM\\
	  	& I5 Evaluates an approach quantifying TD\\
	  	\midrule
	 Exclusion Criteria & \\
	  	& E1 Not a primary study\\
	  	& E2 The paper has not been peer reviewed\\
	  	& E3 The paper is not in English\\
	  	& E4 Research is not directly related to Quantifying TD\\
	  	& E5 Full text is inaccessible\\
\end{tabularx}
\caption{Inclusion and Exclusion Criteria}
\label{tab:inclexcl}
\end{table}

\subsection{Reference Snowballing}

Backward reference snowballing \cite{wohlin2014guidelines} was conducted as an additional step to avoid the possibility of missing articles and was conducted in a semi-automatic manner. First, we exported CSV files of reference lists of articles we already included and then ran a script to check if the list of references had articles that we had already included in our list of primary studies and excluded them. The remaining list was examined to remove any duplicates within the list. After that, the final list of new articles was examined manually to determine if they were relevant or not. Any article resulting from this step was considered
as a candidate for inclusion, the process was iterated over these newly found articles until there were no more candidates for inclusion. 

\subsection{Data Extraction, Data Synthesis and Analysis}

We followed recommendations by Petersen et al. \cite{petersen2008systematic} and Braun and Clarke \cite{braun2006using} for data extraction, synthesis, and analysis. Data extraction was performed on 113 articles that were chosen through the article screening and selection phase in Iteration 1 and then on 16 articles in Iteration 2 of the mapping study. 

We followed the \emph{thematic analysis} approach recommended by Braun and Clarke \cite{braun2006using}, which is an effective method for identifying, analyzing, and reporting patterns and themes within data. We also followed the \emph{adaptive depth reading approach} that we followed in the article screening and selection phase in this phase too. We read the article’s title, abstract, and conclusion, scanning for keywords, and then read, in detail, the sections of the article where we saw relevant information. 

Furthermore, we developed a \emph{classification scheme} following the \emph{keywording approach} recommended by Petersen et al. \cite{petersen2008systematic}. For this, we first classified an initial sample set of studies to extract keywords and then used these keywords to build the classification scheme. We grouped together related keywords and labeled them as a category. Within the categories, we also identified subcategories. The classification scheme was discussed among all authors for consensus. 

In our classification scheme, we categorized the information extracted from the studies into two main categories: Demographics and  Quantification Approaches. Demographics were further categorized into sub-categories, Publication year, and Type of TD. To analyze `Quantification Approaches,' the following information had to be identified; what types of quantification approaches were proposed, what were they based on, what aspects of TD were discussed in the quantification approaches, what software metrics were discussed in them, what units of measurement were discussed in them and what limitations could be found in these approaches. We report on the results of the systematic mapping study in Section \ref{sec:sms_results}.

\section{RQ1: Exploring approaches to TD quantification --- a Systematic Mapping Study --- Results}
\label{sec:sms_results}

A Systematic Mapping Study (SMS) typically reports results firstly in terms of demographics which describe metadata of the primary studies, and secondly, the findings obtained via the understanding of the primary studies. Following the same format, our classification scheme was based on two main categories; Demographics and Quantification Approaches. We report the results for Demographics in section \ref{sec:demographics} and then the findings obtained for Quantification Approaches in Section \ref{sec:findings}. 

\subsection{Demographics}
\label{sec:demographics}

\subsubsection{Publication Year} 
\begin{figure}[h] 
  \centering
  \includegraphics[width=\textwidth]{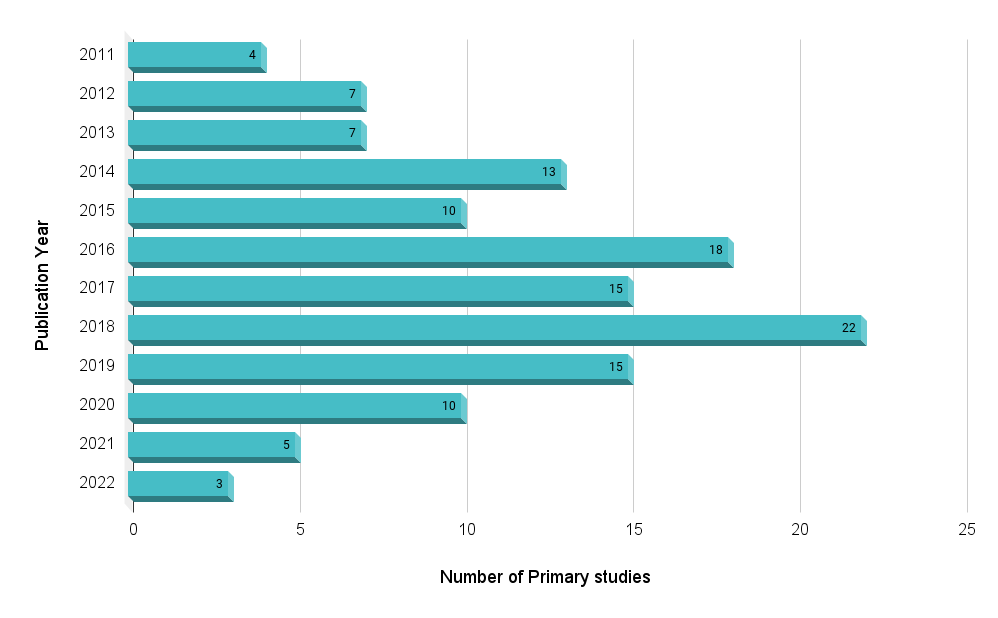}
  \caption{Categorization of primary studies according to Publication Year}
  \label{fig:pubyear}
\end{figure}

We categorized Primary Studies according to the year they were published (See Figure \ref{fig:pubyear}). Publications ranged from 2011 to 2022. The highest number of primary studies related to TD quantification were published in 2018 (22 primary studies). 

\subsubsection{Type of TD} 
\begin{figure}[h] 
  \centering
  \includegraphics[width=\textwidth]{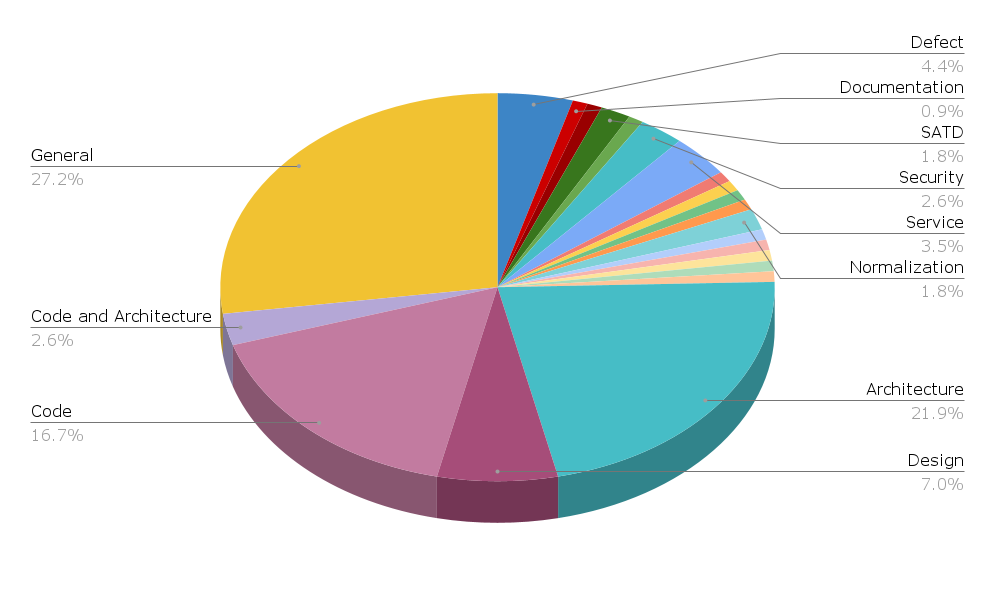}
  \caption{Categorization of primary studies according to type of TD}
  \label{fig:type_of_TD}
\end{figure}

Table \ref{tab:typesoftd} shows the mapping between the number of primary studies and the types of TD discussed in the primary studies. However, some papers did not specify a type of TD. We categorized them under the category ’General’  (most primary studies, 27.2 percent as illustrated in Figure \ref{fig:type_of_TD}, belonged to this category). Architecture (21.9 percent), Code (16.7 percent), and Design TD (7.0 percent) followed as the most popular types of TD among the primary studies found in our mapping study. 

\begin{table}[h]
\centering
\begin{tabularx}{\textwidth}{l|l} 
		\midrule
    \multicolumn{2}{l}\emph{\textbf{SMS Iteration 1 --- Studies from 2011 - 2020}} \emph{--- 113 Primary Studies in Total}\\
		\midrule
	Defect & 4\\
    Documentation & 1\\
    Requirements & 1\\
    SATD & 2\\
    Scalability & 1\\
    Security & 2\\
    Service & 4\\
    Social & 1\\
    Sustainability & 1\\
    Compliance & 1\\
    Elasticity & 1\\
    Normalization & 1\\
    TD in BPMN & 1\\
    TD in ML & 1\\
    TD in Model Transfermation Languages & 1\\
    TD Related to DB Schemas & 1\\
    Temporal Adaption & 1\\
	\emph{Architecture} & 22\\
	\emph{Design} & 7\\
	\emph{Code} & 11\\
	\emph{Code and Architecture} & 3\\
	\emph{General (inferred as code-related)} & 30\\
			\midrule
	\multicolumn{2}{l}\emph{\textbf{SMS Iteration 2 --- Studies from 2020 - 2022}} \emph{--- 16 Primary Studies in Total}\\
		\midrule
	Defect & 1\\
    Security & 1\\
    Normalization & 1\\
	\emph{Architecture} & 3\\
	\emph{Design} & 1\\
	\emph{Code} & 8\\
	\emph{General (inferred as code-related)} & 1\\
	\\
\end{tabularx}
\caption{Number of Primary Studies according to Type of TD | Italicised items --- code related types of TD}
\label{tab:typesoftd}
\end{table}

\subsection{Quantification Approaches} 
\label{Existing Approaches to quantifying TD}
\label{sec:findings}

Multiple approaches to quantifying TD were identified during the mapping study. However, different approaches focus on different aspects of TD. Some approaches base their quantification on identifying code, design, or architectural smells \cite{Sharma2016, Fontana2017, zazworka2011prioritizing, verdecchia2020atdx, roveda2018towards, charalampidou2017assessing, fontana2015towards}. Some approaches try to quantify the Return on Investment (ROI) of refactoring \cite{Kazman2015}. Some approaches compare an ideal state with the current state in terms of software quality \cite{Singh2014, Reimanis2016, Letouzey2016, von2019mitigating, nugroho2011empirical, martini2018semi}. Some approaches compare alternative paths to development to reduce rework \cite{Nord2012, Sangwan2020}. 

However, it is unclear if these different approaches are quantifying the same thing and if they support similar or different decisions regarding TDM. To illustrate this, we describe two approaches, Nord et al. \cite{Nord2012} and Kazman et al. \cite{Kazman2015} in detail in the following two sub-sections and then illustrate the difficulty in comparing these approaches in Section \ref{Sec:Comparing_approaches_NordvsKazman}. The same two approaches are used as examples in Section \ref{example application of TDQM} to illustrate how our model facilitates the systematic comparison of the two approaches. We chose these two approaches because they had a high count of citations and were developed by reputed research groups (107 and 92 citations in SCOPUS , 193 and 151 citations in Google Scholar).

\subsubsection{Nord et al., 2012}

Nord et al. \cite{Nord2012} suggest the comparison of alternative development paths during release planning for the purpose of determining the most suitable path for development. Two development paths are considered in their case study: The path where the developers take on TD and the path where the developers do not take on TD. The total cost of a development path is compared against the total cost of another development path. According to the authors, the total cost is a function of implementation and rework costs. It is calculated by combining the total implementation cost of new architectural elements to be added in a release with the total cost to rework pre-existing elements. 

Although not explicitly stated, the paper implies that ‘rework’ is considered TD, and the end goal is to reduce ‘rework’ by selecting the path which incurs relatively fewer rework costs. The product of, the implementation cost of each pre-existing element, the number of dependencies (direct and indirect) between the new and the pre-existing elements, and the overall Change Propagation (CP) metric of the system (initially introduced by MacCormack et al. \cite{MacCormack2012}) is calculated as the rework cost.

\subsubsection{Kazman et al., 2015}

Kazman et al. \cite{Kazman2015} discuss the repayment of Architectural Technical Debt (ATD) in their paper. They calculate the Return on Investment (ROI) of refactoring the architectural issues pertaining to the debt. The authors use the `Design Rule Space analysis' approach \cite{xiao2014design} to identify the Design Rule Spaces \footnotemark[1]{}\footnotetext[1]{A form of architecture representation that uniformly captures both architecture and evolution structures. \cite{xiao2014design}} that capture the software system's most error-prone and change-prone files. Then from those architectural roots, architecture issues are further diagnosed and extracted as architecture debts. 

Their quantification framework consists of two main functions: the first \emph{calculates the penalty incurred by the DRSpaces} and the second \emph{estimates the benefit that can be expected to be accrued by refactoring the DRSpaces}. After that, the expected benefits are compared with the cost of refactoring to determine the Return on Investment (ROI). 

In their case study, effort proxies were taken from the revision history and issue tracking systems to measure the penalty associated with the debt. Examples include; the number of resolved defects per file, the number of completed changes per file, and the number of modified or added or deleted lines of code per file to fix defects and make changes. The refactoring cost was estimated in Person Months based on expert opinion.

The expected benefit from refactoring is the difference between the actual yearly numbers of defects, changes, and committed LOC and the expected numbers of defects, changes, and committed LOC after refactoring. For example, the benefit in terms of LOC, i.e., the LOC expected that the project would not have to commit in the future due to doing the refactoring, is \emph{the difference between the actual (current) number of LOC and the expected number of LOC after refactoring per DRSpace per year}. The company's average productivity (e.g., 600 LOC per month) value is then used to calculate the expected Person Months (PM) avoided or saved as the ROI of refactoring. 

\subsection{Comparing and Evaluating Quantification Approaches}
\label{Sec:Comparing_approaches_NordvsKazman}

Due to the wide variety of concepts discussed in the quantification approaches found in the mapping study, we found it difficult to compare and evaluate them. We illustrate the difficulty faced in comparing quantification approaches using the two approaches, Nord et al. \cite{Nord2012} and Kazman et al. \cite{Kazman2015} in Table \ref{tab:comparisondifficultyexample}. Due to space limits, we do not attempt to list all the concepts extracted from all quantification approaches in this paper. They can be found in our Replication Package  \footnotemark[2]{}\footnotetext[2]{Replication Package: \url{https://drive.google.com/drive/folders/1rcCgupChUkET9KGntlBNQ9-o-EF-hIsL?usp=share_link}}. 

Referring to Table \ref{tab:comparisondifficultyexample}, although we could extract concepts related to TD quantification from both approaches Nord et al. \cite{Nord2012} and Kazman et al. \cite{Kazman2015}, we could not easily create a mapping between the two lists of concepts. It is difficult to determine if both approaches are quantifying the same by simply looking at the two lists of concepts. It is also difficult to determine if they support similar or different decisions regarding TDM decision-making. One may somehow try to understand and compare the two approaches. However, it takes much work to do so. It becomes more difficult to compare multiple approaches as opposed to two approaches as it is inefficient to compare them all by pair-wise comparisons between them. This inefficiency highlights the need for a method for systematic comparison and evaluation of quantification approaches, the motivation to develop our model, a unified common model that allows systematic comparisons and evaluations between quantification approaches. Section \ref{sec:method:tdqm} describes the methodology for the development of the model. Section \ref{sec:results:tdqm} presents the model. Section \ref{sec:method:tdqm_app} and \ref{sec:results:tdqm_app} report on the methodology and results of applying the model to quantification approaches found in the mapping study before and after the development of the model.

\begin{table}[htb]
\centering
\begin{tabularx}{\textwidth}{l|l} 
	\textbf{Concepts extracted from NOK+12} & \textbf{Concepts extracted from KCM+15} \\
		  	\midrule
	Product & Refactoring Step\\
    Release & Architectural Flaw\\
	Development Path & Cost of Refactoring\\
	Feature & Penalty incurred by Debts\\
    Cumulative Total Cost & Expected benefit of Refactoring\\
    Percentage of Cumulative Total Cost & \\
    Implementation Cost & \\
    Rework Cost & \\
    \\
\end{tabularx}
\caption{Concepts extracted from two quantification approaches}
\label{tab:comparisondifficultyexample}
\end{table}

\section{RQ2: Modelling the quantification of TD --- the development of a Conceptual Model --- Methodology}
\label{sec:method:tdqm}

The necessity of having a systematic method to compare and evaluate different quantification approaches (discussed in Section \ref{sec:sms_results}) motivated the development of the Technical Debt Quantification Model (TDQM). Through the development of TDQM, we answered the following Research Question, RQ2.

\begin{itemize}
    \item \emph{\textbf{RQ2: How can we model TD quantification?}}
    \begin{itemize}
        \item \emph{\textbf{RQ2.1: What are the concepts that pertain to TD Quantification?}}
        \item  \emph{\textbf{RQ2.2: What are the relationships that can be identified among such concepts?}}
    \end{itemize}
\end{itemize}

\begin{figure*}
  \centering
  \includegraphics[width=\textwidth]{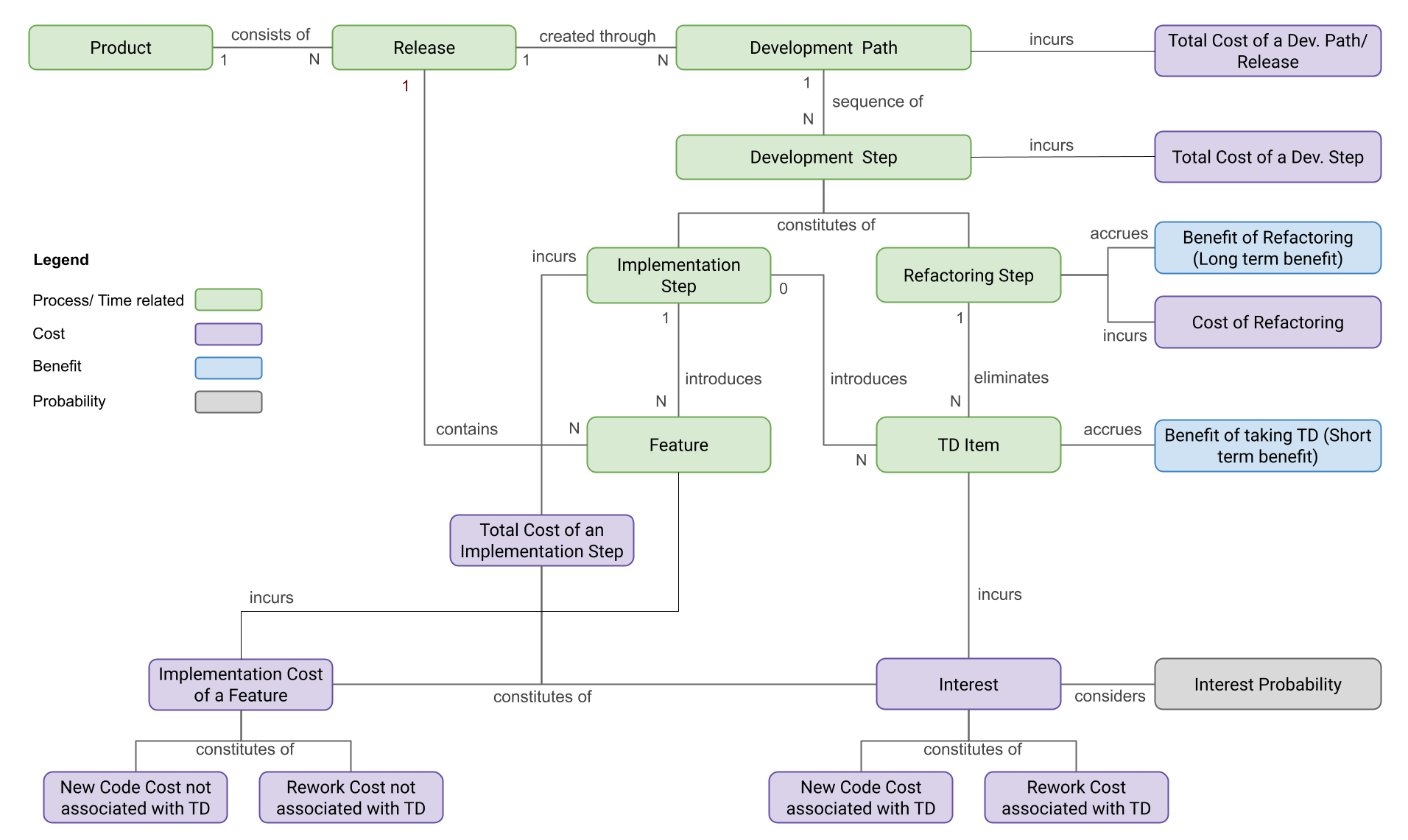}
  \caption{TDQM Concept Map}
  \label{fig:Concept_Map_model}
\end{figure*}

We developed the \emph{Technical Debt Quantification Model (TDQM)} (See Concept Map in Figure \ref{fig:Concept_Map_model}), in part, by examining what constitutes TD quantification and in part, by examining how software development can be affected by TD. We observed the need to represent the introduction of TD, the removal of TD, and how the consequences of leaving the TD Items over time could affect software development. TDQM was also informed by past literature and by past models of TD. For example, the concept of a \emph{`TD Item'} was informed by the \emph{16162 model} \cite{avgeriou2016managing} described in Section \ref{sec:related_work}. 

We examined a sample set of primary studies found in our mapping study and then extracted various aspects related to TD quantification to capture the information that could answer RQ2. We identified commonly discussed themes (e.g., process, time, costs, benefits, probability), concepts (e.g., TD Interest, refactoring Cost) and relationships between them. This helped us understand how the concepts fit together and how it aligns with our experience in software development in the industry. By that, we could answer RQ 2.1 and RQ2.2. 

The notions of \textbf{time, cost, and benefit} were identified as recurrent themes in the sample set of quantification approaches. Therefore, we considered them important aspects to represent in our model. For example, the TDQM concept \emph{`Development Step'} captures time. \emph{`Cost of Refactoring'} represents a cost while \emph{`Benefit of Refactoring'} represents a benefit. Ribeiro et al. introduce criteria that can be utilized for TDM decision-making. They identified \emph{`Debt impact on the project’} and \emph{`Cost-Benefit’} as the most explored criteria in the studies captured in their mapping study [20]. This confirms our selection of TDQM concepts (i.e., concepts related to cost and benefit, including TD Interest) that we chose to build our theory.

The concept of \emph{`Rework'} was first discussed in Nord et al.'s  \cite{Nord2012} approach. Further exploring the concept, we observed that rework, in some cases, could be associated with TD and, in some cases, could also not be associated with TD. Hence, in our model, we illustrate that rework does not necessarily associate with only TD; developers may rework to enhance the code without necessarily having TD Items present in that code. Hence, in our model, we decompose `rework cost' into \emph{`Rework cost associated with TD} and \emph{`Rework Cost not associated with TD'}. Similarly, we observed that \emph{`New Code Cost'} can have its TD-associated and non-associated counterparts. Developers may write new code to implement new features, while they may also write new code as a workaround not to remove the TD when TD is present.

TDQM was developed in iterations by following the above process of extracting concepts from a subset of quantification approaches from our mapping study, identifying the relationships between these concepts, and discussing them among the researchers before including them in the model. At some point, no new concepts were identified apart from what we could not categorize as the commonly identified model concepts. We then applied the model to another subset of quantification approaches to determine its feasibility. Since the model was applicable, it was deemed to be comprehensive enough to model the quantification of TD. 

We developed TDQM to model the quantification of TD types related to software code, for example, Code, Design, Architectural, and General (inferred as code-related) types of TD. The reasoning behind this was that most static analysis tools used in the \emph{identification of TD} focus on code-related TD types of TD e.g., SonarQube\footnotemark[3]{} \footnotetext[3]{https://www.sonarqube.org/}, DV8 \cite{cai2019dv8}, Designite \cite{Sharma2016}. Therefore, we considered this a starting point for modeling the quantification of TD. We plan to extend the model to other types of TD in our future studies.

\section{RQ2: Modelling the quantification of TD --- the development of a Conceptual Model --- Results}
\label{sec:results:tdqm}

\subsection{Technical Debt Quantification Model (TDQM)}

The Technical Debt Quantification Model (TDQM) captures the important concepts related to TD Quantification and illustrates the relationships between these concepts (See Figure \ref{fig:Concept_Map_model}). TDQM serves as a common model that enables effective comparisons and evaluations among quantification approaches as well as serves as a reference point to develop new quantification approaches.

\subsubsection{Modelling Development of a Product}

We model the development of a software \emph{Product} (See Figure \ref{fig:Concept_Map_model}) as consisting of a sequence of \emph{Releases}. Each release contains a (usually different) set of \emph{Features}. A release is created through a \emph{Development Path}. Depending on the development team's choices, there may be different development paths for a given release. For example, in one path, the team may take on more TD than in another path. A specific development path is a sequence of \emph{Development Steps}. A \emph{Development Step} may consist of the addition of features to the product (which we call an \emph{Implementation Step}) or the removal of TD Items (\emph{Refactoring Step}). An \emph{Implementation Step} may or may not introduce TD (represented by the 0 to N relationship in Figure \ref{fig:Concept_Map_model}). 

Following the 16162 model \cite{avgeriou2016managing}, we model TD as \emph{TD Items}. TD could be introduced deliberately or inadvertently during the implementation of a feature. Furthermore, the removal of TD can be based on \emph{Prioritization}, another TDM activity \cite{lenarduzzi2021systematic}. There could be other development activities, such as bug fixing or feature enhancements. However, we keep the model simple by modeling the development activities sufficient to model the quantification of TD.

\subsubsection{Modelling Development Path Cost, Development Step Cost, Implementation Step Cost and Cost of Refactoring}

A development path that leads to a release of a product incurs a cost (i.e., time and effort spent to develop the product by following that path), the \emph{Total Cost of a Development Path}. This cost will be calculated by adding up the total costs of the development steps. Each development step will incur a total cost, \emph{the Total Cost of a Development Step}, which will either be the \emph{Total Cost of an Implementation Step} which consists of the \emph{Implementation Cost of a Feature} and the TD \emph{Interest} (if there is TD) if the \emph{Development Step} is an \emph{Implementation Step}, or \emph{Cost of Refactoring} if the development step is a \emph{Refactoring Step}. 

\subsubsection{Modelling Interest and Interest Probability}

The \emph{extra or additional cost that is the consequence of the presence of TD (i.e., TD Interest)} incurs depending on how much TD is present at the time of the implementation. The additional cost is the sum of the Interest for each TD Item. Although, the generation of Interest for a TD Item also depends on if the implementation interacts with a TD Item. For example, a TD Item may be entirely enclosed in one class, and the implementation does not require any reference to that class. Whether or not implementation is affected by a TD Item is referred to as the \emph{Interest Probability} \cite{seaman2012using}. The impact of a TD Item being introduced during an \emph{Implementation Step} occurs only on the \emph{next} step (and all future steps until it is removed) and not on the step where it is being introduced.

\subsubsection{Modelling cost components --- New Code Cost and Rework Cost}

We further decompose the \emph{Implementation cost of a feature} and the \emph{Interest} into constituents; \emph{New Code Cost} and \emph{Rework Cost}. We saw the need to do so to model the costs that will be incurred due to writing new code and doing rework as described in Section \ref{sec:method:tdqm}. Any implementation step requires writing the \emph{new code} for the feature being added --- this has a cost (\emph{New Code Cost not Associated with TD}). However, some of the existing code must also change, at minimum, to integrate the new code, and this \emph{rework} also has a cost (\emph{Rework Cost not Associated with TD}). TD Interest adds to the total cost of an implementation step in a similar way. Some new code may need to be written to work around the issues associated with the TD (\emph{New Code Cost associated with TD}), and some rework of existing code may be needed as well (\emph{Rework Cost associated with TD}).

\subsubsection{Modelling Benefit of Refactoring and Benefit of taking TD}

Benefits accrued during a development of a product could be either short-term or long-term. The short-term benefit gained is the benefit accrued by taking TD (\emph{Benefit of taking TD}), utilizing TD strategically, while the long-term benefit would be the \emph{Benefit of Refactoring}, accrued as a benefit in the long run by reestablishing code quality in the long run. The short-term benefit allows faster delivery to market at the beginning of the project while the long-term benefit reduces the time to add features to a product in the future. 

\subsection{Application of TDQM}
\label{Application of TDQM} 

In this Section, we demonstrate the application of TDQM, that is, how it is utilized to compare and evaluate various quantification approaches.  
 
\subsubsection{Data Extraction}
\label{process: data extraction}

The first step in applying TDQM to compare and evaluate quantification approaches is to identify and extract the process/time, cost, benefit, and probability-related concepts from the approaches. For this, we traverse the TDQM concepts using it as a tool to extract concepts from the individual quantification approaches while determining if the approach concepts correspond to the concepts in the TDQM Concept Map (Figure \ref{fig:Concept_Map_model}). 

\subsubsection{Mapping to TDQM Concepts}
\label{process: mapping}

\begin{figure*}
  \centering
  \includegraphics[width=\textwidth]{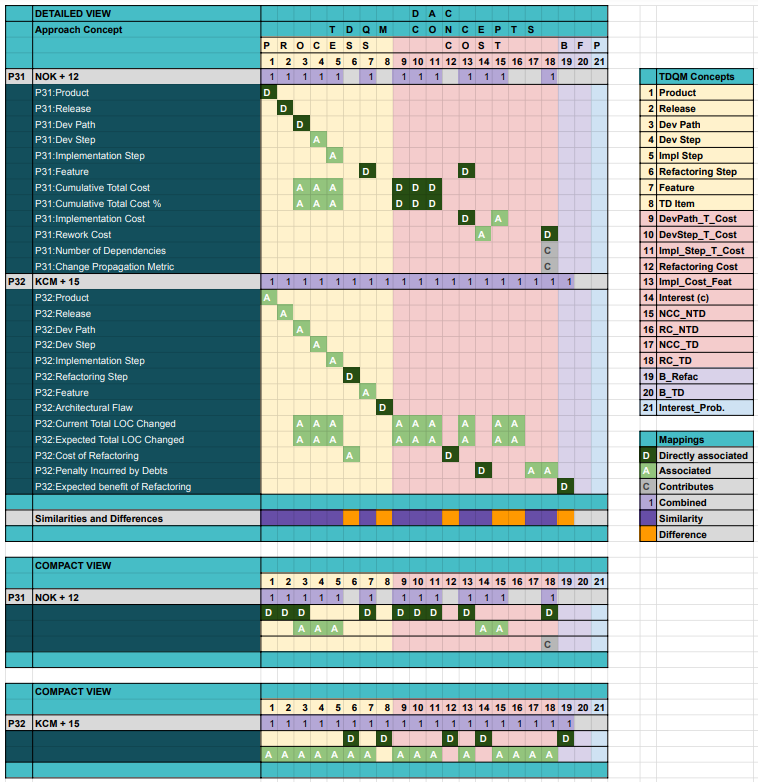}
  \caption{Mapping of the two quantification approaches, Nord et al. (NOK + 12) and Kazman et al. (KCM + 15), to TDQM Concepts | Top: Detailed View visualizing individual mappings between TDQM concepts and the approach concepts, Bottom: Compact View visualizing a compact version of the mappings \emph{D, A} and \emph{C}} 
  \label{fig:viz comparison}
\end{figure*}

In the second step, we map TDQM concepts and the concepts extracted from the quantification approach. Figure \ref{fig:viz comparison} illustrates the mapping of the concepts of the two example quantification approaches to the concepts of TDQM, which we discuss further in Section \ref{example application of TDQM}. We indicate the \emph{level of association} between the concepts of TDQM and the concepts found in the quantification approaches by the terms \emph{D, A} and \emph{C} (defined below). D and A are on the same dimension as they describe \emph{`the degree to which a given approach concept maps to a TDQM concept'}. C pertains to a different dimension; it indicates the mapping between \emph{metrics} identified in the quantification approach to the TDQM concepts.

\begin{itemize}
    \item \textbf{`Direct' mapping --- indicated by \emph{(D)}}: The approach concept corresponds exactly to the TDQM concept and it is straightforward to extract.
    \item \textbf{`Associated' mapping --- indicated by \emph{(A)}}: The approach concept relates to the TDQM concept in some way but does not correspond exactly. i.e., it might be inferred, not straightforward to extract.
    \item \textbf{`Contributes' mapping --- indicated by \emph{(C)}}: A metric discussed in the approach which contributes to the calculation of the TDQM concept.
    \item \textbf{`Combined' mapping --- indicated by \emph{(1)}} --- Combines mappings \emph{D} and \emph{A} into a single mapping in order to provide a simpler view. 
\end{itemize}

\subsubsection{Comparison and Evaluation}
\label{process: comparison and evaluation}

The third step in the process involves comparing and evaluating the quantification approaches against each other using the mappings done in step \ref{process: mapping}. As stated before, the problem we are trying to solve via TDQM is to be able to compare and evaluate various quantification approaches usefully and efficiently. In particular, we try to determine what precisely an approach quantifies in terms of TDQM concepts and by that, if the approach can be efficiently compared against other approaches to determine if they are quantifying the same thing. 

TDQM allows representing the two example quantification approaches in a common format via the \emph{TDQM Approach Comparison Matrix} (See Figure \ref{fig:viz comparison}, described in Section \ref{example application of TDQM}) that allows seeing what TDQM concepts are covered and not covered by the two quantification approaches, at the same time. The TDQM Approach Comparison Matrix increases the efficiency in performing comparisons and evaluations between the two quantification approaches.  

\subsection{Example Application of TDQM --- Comparing and Evaluating two Approaches}
\label{example application of TDQM}

Here, we demonstrate the application of TDQM to compare and evaluate quantification approaches. We use two example quantification approaches from the research literature (Nord et al. \cite{Nord2012}, and Kazman et al. \cite{Kazman2015}, which were discussed in Section \ref{Existing Approaches to quantifying TD}). In particular, we answer the question \textbf{"What do the proposed approaches quantify?"} for the two approaches, with the support of TDQM.

Figure \ref{fig:viz comparison} illustrates the representation of the two example quantification approaches in what we call the \emph{\textbf{`TDQM Approach Comparison Matrix'}}. The matrix allows for performing efficient comparisons and evaluations between the quantification approaches. It categorizes the TDQM concepts into \emph{process/time, costs, benefits and probability}, similar to the categorization in the TDQM Concept Map (Figure \ref{fig:Concept_Map_model}). The matrix comes in two views; the \emph{\textbf{Detailed View}} and the \emph{\textbf{Compact View}}.

The top block in Figure \ref{fig:viz comparison} displays the \emph{Detailed View} of the matrix, visualizing individual mappings between TDQM concepts and the approach concepts, while the bottom block displays the \emph{Compact View} of the matrix, visualizing a compact version of the mappings D, A and C. In the \emph{Detailed View}, the approach concepts are listed in the \emph{\textbf{rows}} prefixed with \emph{P31} and \emph{P32} for Nord et al. \cite{Nord2012} and Kazman et al. \cite{Kazman2015} respectively. Each of them is mapped individually to the TDQM Concepts in the \emph{\textbf{columns}} denoted 1-21 following the labeling described in Section \ref{process: mapping}. A combined version of the mappings, which combines the mappings for \emph{D} and \emph{A} both into one mapping, is displayed as the header row for each approach in both \emph{Detailed} and \emph{Compact} views.

Now that we have represented both quantification approaches, Nord et al. and Kazman et al. via the \emph{TDQM Comparison Matrix} which is a uniform representation, we can use it to perform a systematic comparison between the two approaches.

We can see that the approaches are similar in some aspects, while they also differ in what they quantify. Both approaches \textbf{commonly quantify}, for example, \textbf{the concepts}, Total cost of a development path (9), Total cost of a development step (10), Total cost of an Implementation (11), Implementation cost of a feature (13), Interest (14), New Code cost associated with TD (17) and Rework cost associated with TD (18) while we can also see that the two approaches have \textbf{some concepts that are not in common}, for example, TD Item (8), Refactoring cost (12), New Code Cost not associated with TD (15), Rework cost not associated with TD (16) and Benefit of Refactoring (19), with respect to the concepts of TDQM as per illustrated in Figure \ref{fig:viz comparison}. Here we made the comparison based on the combined mapping regardless of the more detailed mappings \emph{D and A}. Following is a detailed comparison and evaluation of the two approaches, taking into account the more detailed mappings, i.e., level of association \emph{D and A}.

Both approaches cover the TDQM concepts; Product (1), Release (2), Development Path (3), Development Step (4), Implementation Step (5), Feature (7), Total cost of a Development Path (9), Total cost of a Development Step (10), Total cost of an Implementation Step (11), Implementation cost of a Feature (13), Interest (14), New Code cost associated with TD (17) and Rework cost associated with TD (18). \textbf{Yet, the degree of association might be different}.
Consider the TDQM Concept Total cost of a Development Path (9) as an example. Although Nord et al.'s approach has a `Cumulative Total Cost' concept that is directly associated (D) with the TDQM concept, Kazman et al. have only an associated (A) concept, `Current Total LOC Changed,' for the same TDQM concept. \emph{D} can be considered the stronger mapping compared to \emph{A}. However, the TDQM representation has allowed us to identify that Kazman et al. quantify the same concept even though it is not explicitly mentioned in their paper.

Nord et al. do not discuss Refactoring and do not quantify either the Refactoring cost (12) or the benefit of Refactoring (19). In comparison, these concepts are being quantified by Kazman et al. Kazman et al.'s `Cost of Refactoring' corresponds to the TDQM concept Refactoring Cost (12), which also implies the associated mapping (A) with Refactoring Step. `Expected Benefit of Refactoring' directly maps (D) to the TDQM concept Benefit of Refactoring. 
Nord et al. do not refer to a TD Item (8) concept. Hence, we cannot determine the addition or removal of TD items. However, Kazman et al. identify TD as 'Architectural Flaws,' which can be mapped directly (D) to the concept of a TD Item in TDQM. 

Nord et al. explicitly mention the quantification of `Cumulative Total Cost,' which can be directly mapped (D) to the Total Cost of a Development Path (9). Nord et al. also explicitly mention the Total Cost of a Development Step (10) and `Implementation Cost' that can be directly mapped (D) to the Total Cost of an Implementation Step (11) and Implementation Cost of a Feature (13). At the same time, Kazman et al. do not quantify these concepts explicitly. However, considering that the states before and after refactoring are corresponding to the two TDQM variants of implementation, \emph{introducing TD} and \emph{not introducing TD} during an implementation step (denoted by the $0:N$ relationship in Figure \ref{fig:Concept_Map_model}), also considering that a development path consists of a single development step in Kazaman et al.'s case study, we can infer that these costs are being quantified (A). 

Nord et al. directly (D) quantify the Implementation cost of a Feature (13) as `Implementation Cost.' We could interpret it as quantifying the constituent New Code cost not associated with TD (15) but not the constituent Rework cost not associated with TD (16) concerning TDQM concepts. Their 'Rework Cost' is defined as TD Interest. Hence, it refers only to the constituent Rework cost associated with TD. The quantification of the implementation cost of a feature is not straightforward in Kazman et al.'s approach. Although, if we assume that the files being modified belong to the same feature, we can consider this concept as quantified in their approach. Since they refer to LOC committed to making changes, we can consider that they quantify the Implementation cost of a Feature in terms of both New code cost not associated with TD (15) and Rework cost not associated with TD (16). This is illustrated by the Associated mappings (A) to ``Current Total LOC Changed'' and `Expected Total LOC Changed.'

The interest in Nord et al. corresponds to only the Rework cost associated with TD (as discussed in the previous paragraph). At the same time, Kazman et al. quantify both constituents of interest, New code cost associated with TD (17) and Rework cost associated with TD (18), as the `penalty (or interest) incurred by the debts.' This is evident since they refer to the LOC committed to making changes. 

However, the TDQM Comparison Matrix visualizes that both approaches do not facilitate quantifying the TDQM concepts; the Benefit of taking TD (20) (i.e., the short-term benefit realized in a software project) and the concept 'Interest Probability' (21) (i.e., the probability of incurring interest).

\section{RQ3: Comparing and evaluating quantification approaches --- a study using TDQM --- Methodology}
\label{sec:method:tdqm_app}

We used TDQM in a study to compare and evaluate quantification approaches found in our mapping study prior to and after the development of TDQM. The research question we answered via this study is;

\begin{itemize}
    \item \emph{\textbf{"RQ3: How can we compare and evaluate quantification approaches?
"}}.
\end{itemize}

As described previously in Section \ref{sec:method:tdqm} (TDQM Methodology), TDQM was developed for modeling the quantification of code-related types of TD such as, Code, Design, Architectural, and General (inferred as code-related). Therefore, TDQM was applied to quantification approaches of the same types of TD. Table \ref{tab:typesoftd_SMS_it} illustrates the derivation of \emph{unique quantification approaches for code related types of TD} from both iterations of the systematic mapping study; Iteration 1 and 2. Although Iterations 1 and 2 initially obtained 73 and 13 primary studies for code-related types of TD after grouping similar or related case studies, the final results of the derivation included 33 and 6 primary studies for Iterations 1 and 2, respectively. This resulted in 39 unique quantification approaches from both iterations of the mapping study. In Section \ref{sec:results:tdqm_app}, we report the results of applying TDQM to compare and evaluate these 39 quantification approaches.

Firstly, we followed the process described in Section \ref{Application of TDQM} to map the quantification approaches to TDQM and to represent them via the \emph{TDQM Approach Comparison Matrix}. Then we utilize different sortings of the TDQM Approach Comparison Matrix in its \emph{Compact View} that we initially discussed in Figure \ref{fig:viz comparison} and in the example in Section \ref{sec:results:tdqm}, to visualize different classifications of the 39 quantification approaches. Representing the quantification approaches using the TDQM Approach Comparison Matrix helps better understand the approaches with respect to TDQM Concepts. Thereby, it becomes possible to efficiently perform comparisons and evaluations between the various quantification approaches. The approach comparison matrix also allows identifying trends among the various quantification approaches, for example, Publication Year Vs. TDQM Concepts, and TD Type Vs. TDQM Concepts. 

\begin{table}[!htb]
\centering
\begin{tabularx}{\textwidth}{l|l} 
		\midrule
        \multicolumn{2}{l}\emph{\textbf{SMS Iteration 1 --- Studies from 2011 - 2020}}\\
		\midrule
	Defect & 4\\
    Documentation & 1\\
    Requirements & 1\\
    SATD & 2\\
    Scalability & 1\\
    Security & 2\\
    Service & 4\\
    Social & 1\\
    Sustainability & 1\\
    Compliance & 1\\
    Elasticity & 1\\
    Normalization & 1\\
    TD in BPMN & 1\\
    TD in ML & 1\\
    TD in Model Transfermation Languages & 1\\
    TD Related to DB Schemas & 1\\
    Temporal Adaption & 1\\
		\midrule
	\emph{Architecture} & 22\\
	\emph{Design} & 7\\
	\emph{Code} & 11\\
	\emph{Code and Architecture} & 3\\
	\emph{General} & 30\\
		\midrule
	\emph{Total} num of Studies for \emph{All types of TD} & 113\\
	    \midrule
	\emph{Total} num of Studies for \emph{\textbf{Code related types of TD}} & \textbf{73}\\
		\midrule
	\emph{\textbf{Unique quantification approaches}} 
	\\for \emph{Code related types of TD} & \textbf{33}\\
		\midrule
	    \multicolumn{2}{l}\emph{\textbf{SMS Iteration 2 --- Studies from 2020 - 2022}}\\
		\midrule
	Defect & 1\\
    Security & 1\\
    Normalization & 1\\
		\midrule
	\emph{Architecture} & 3\\
	\emph{Design} & 1\\
	\emph{Code} & 8\\
	\emph{General} & 1\\
		\midrule
	\emph{Total} num of Studies for \emph{All types of TD} & 16\\
	    \midrule
	\emph{Total} num of Studies for \emph{\textbf{Code related types of TD}} & \textbf{13}\\
		\midrule
	\emph{\textbf{Unique quantification approaches}} 
	\\for \emph{Code related types of TD} & \textbf{6}\\
			\midrule
	\emph{\textbf{Total Unique quantification approaches from both SMS Iterations}}
	\\for \emph{Code related types of TD} & \textbf{${33 + 6 = 39}$} \\
	\\
\end{tabularx}
\caption{Derivation of Unique Quantification Approaches for \emph{Code related types of TD}}
\label{tab:typesoftd_SMS_it}
\end{table}

However, we report only a limited amount of results in this paper. We provide a Replication Package\footnotemark[2] where all the results and the derivation of results can be found. The results also theoretically validated TDQM as it is evident that the model applies to the quantification approaches for code-related TD types, found in our mapping study from both iterations conducted before and after the development of the model and the model did not change after the addition of the new approaches. 


\section{RQ3: Comparing and evaluating quantification approaches --- a study using TDQM --- Results}
\label{sec:results:tdqm_app}

\subsection{Direct (D) Mappings to TDQM}
\subsubsection{Direct (D) Mappings to TDQM sorted by TD Type and Pub Year}

\begin{figure*}[htb]
  \centering
  \includegraphics[width=\textwidth]{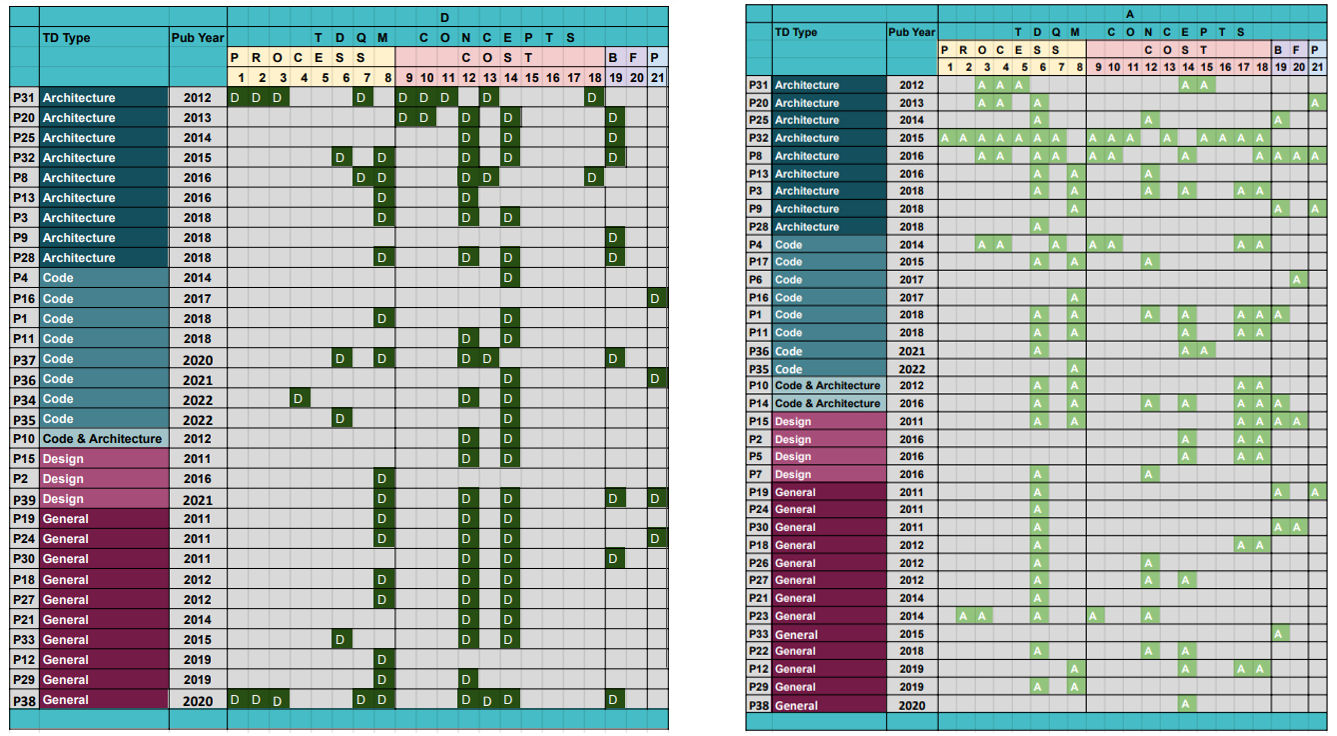}
  \caption{Left: TDQM Approach Comparison Matrix - Compact View : \emph{Direct (D)} Mappings sorted by TD Type and Pub Year | Right: TDQM Approach Comparison Matrix - Compact View : \emph{Associated (A)} Mappings sorted by TD Type and Pub Year}
  \label{fig:1DvsA}
\end{figure*}

In this representation of the TDQM Approach Comparison Matrix (See Left, Figure \ref{fig:1DvsA}), we can see the approaches that have \emph{Direct D)} mappings to TDQM sorted first by the TD Type and then by the Publication Year. Among the approaches that have \emph{D} mappings, nine approaches (2012-2018) quantify Architectural Debt; eight approaches quantify Code TD (2014-2022), and one approach quantifies both Code and Architecture TD (2012). Also, three approaches quantify Design TD (2011-2021), while ten approaches did not specifically mention a type of TD which we categorized as `General'(2011-2020). 

Inside the block of `Architectural TD,' we can observe that `Refactoring Cost' is the most popular concept with seven \emph{D} mappings. For Code TD, it is `TD Interest' with six \emph{D} mappings. For Design TD, `TD Item', `Refactoring Cost' and `Interest' have two \emph{D} mappings each. For the `General' category, it is again `Refactoring Cost,' which is the most popular concept with nine \emph{D} mappings. 

\subsubsection{Direct (D) Mappings to TDQM sorted by Pub Year and TD Type}

In this slightly different representation of the TDQM Approach Comparison Matrix (See Left, Figure \ref{fig:2DvsA}), approaches are sorted first based on the Publication Year and then the TD Type. It is apparent that `Refactoring Cost' and `Interest' were popular discussion topics starting from the early years of TD quantification, 2011 and 2012. `Interest Probability' has been a far less discussed topic than many of the TDQM Concepts. Mappings to this concept can be observed only in 2011, 2017, and 2021. `Rework Cost associated with TD' received even lesser attention and has been discussed only in two studies (P8, P31 | P - Primary Study) in 2012 and 2016, respectively (See Table \ref{tab:primarystud_sms} for the list of primary studies for code-related types of TD found in our mapping study). The concept of `TD Item' has generally been popular throughout. 

\begin{table}[!htb]
\centering
\begin{tabular}{p{1.15cm}|p{12cm}}
Study & Title\\ 
       \midrule
P1 \cite{tornhill2018assessing} & Assessing Technical Debt in Automated Tests with CodeScene\\
P2 \cite{Sharma2016} & Designite - A Software Design Quality Assessment Tool\\
P3 \cite{Martini2018} & AnaconDebt: A Tool to Assess and Track Technical Debt\\
P4 \cite{singh2014framework} & A framework for estimating interest on technical debt by monitoring developer activity related to code comprehension\\
P5 \cite{Reimanis2016} & Towards assessing the Technical Debt of Undesired Software Behaviors in Design Patterns\\
P6 \cite{kosti2017technical} & Technical Debt Principal Assessment through Structural Metrics\\
P7 \cite{mensah2016rework} & Rework Estimation of Self-admitted Technical Debt\\
P8 \cite{vathsavayi2016technical} & Technical Debt Management with Genetic Algorithms\\
P9 \cite{roveda2018towards} & Towards an Architectural Debt Index\\
P10 \cite{curtis2012estimating} & Estimating the Size, Cost, and Types of Technical Debt\\
P11 \cite{ampatzoglou2018framework} & A Framework for Managing Interest in Technical Debt: An Industrial Validation\\
P12 \cite{perez2019proposed} & A Proposed Model-Driven Approach to Manage Architectural Technical Debt Life Cycle\\
P13 \cite{shapochka2016practical} & Practical Technical Debt Discovery by Matching Patterns in Assessment Graph\\
P14 \cite{choudhary2016minimizing} & Minimizing Refactoring Effort through Prioritization of Classes based on Historical, Architectural and Code Smell Information\\
P15 \cite{zazworka2011prioritizing} & Prioritizing Design Debt Investment Opportunities\\
P16 \cite{charalampidou2017assessing} & Assessing Code Smell Interest Probability: A Case Study\\
P17 \cite{fontana2015towards} & Towards a prioritization of code debt: A code smell Intensity Index\\
P18 \cite{letouzey2012managing} & Managing Technical Debt with the SQALE Method\\
P19 \cite{guo2011portfolio} & A portfolio approach to technical debt management\\
P20 \cite{schmid2013formal} & A  Formal Approach to Technical Debt Decision Making\\
P21 \cite{mayr2014benchmarking} & A Benchmarking-based model for Technical Debt Calculation\\
P22 \cite{snipes2018proposed} & A proposed sizing model for managing 3rd party Code Technical Debt\\
P23 \cite{ho2014release} & When-to-release decisions in consideration of TD\\
P24 \cite{guo2011tracking} & Tracking technical debt - An exploratory case study\\
P25 \cite{fernandez2014guiding} & Guilding Flexibility Investment in Agile Architecting\\
P26 \cite{eisenberg2012threshold} & A Threshold based approach to TD\\
P27 \cite{curtis2012estimating} & Estimating the Principal of an Application's TD\\
P28 \cite{martini2018semi} & A semi-automated framework for the identification and estimation of Architectural Technical Debt: A comparative case-study on the modularization of a software component\\
P29 \cite{al2019evolution} & Evolution of TD: An exploration Study\\
P30 \cite{nugroho2011empirical} & An Empirical Model of Technical Debt and Interest (SIG method)\\
P31 \cite{Nord2012} & In Search of Metric for Managing Architectural Technical Debt\\
P32 \cite{Kazman2015} & A Case Study in Locating the Architectural Roots
of Technical Debt\\
P33 \cite{falessi2015towards} & Towards an open-source tool for measuring and visualizing the interest of TD\\
P34 \cite{vora2022measuring} & Measuring the Technical Debt\\
P35 \cite{kontsevoi2022practice} & Practice of Tech Debt Assessment and Management with TETRA\\
P36 \cite{digkas2021risk} & The Risk of Generating Technical Debt Interest: A Case Study\\
P37 \cite{arif2020refactoring} & Refactoring of Code to Remove Technical Debt and Reduce Maintenance Effort\\
P38 \cite{stochel2020continuous} & Continuous Debt Valuation Approach (CoDVA) for Technical Debt Prioritization\\
P39 \cite{nikolaidis2021experience} & Experience With Managing Technical Debt in Scientific Software Development Using the EXA2PRO Framework\\
\\
\end{tabular}
\caption{Primary Studies found in the SMS for code-related types of TD | Pn [m] --- Primary Study [Citation]}
\label{tab:primarystud_sms}
\end{table}

\subsection{Associated (A) Mappings to TDQM}

\subsubsection{Associated (A) Mappings to TDQM sorted by TD Type and Pub Year}

In this representation of the TDQM Approach Comparison Matrix (See Right, Figure \ref{fig:1DvsA}), we can see the approaches that have \emph{Associated A)} mappings to TDQM sorted first by the TD Type and then by the Publication Year. Among the approaches with \emph{A} mappings, nine quantify Architecture TD (2012-2018). Eight approaches quantify Code TD (2014-2018), and two quantify Code and Architecture TD (2012, 2016). Four approaches that have \emph{D} mappings quantify Design TD (2011-2016), while thirteen did not mention the type of TD, which we categorized as `General'(2011-2020). 

Inside the block of `Architectural TD,' we can observe that `Refactoring Step' is the most popular concept with seven \emph{A} mappings to TDQM. For Code TD, it is `TD Item' with five \emph{A} mappings that is the most popular concept. Inside the block of Design TD, `New Code Cost associated with TD' and `Rework Cost associated TD' have the highest number of mappings which is three. For the `General' category, it is again `Refactoring Step' with ten \emph{A} mappings. 

\subsubsection{Associated (A) Mappings to TDQM sorted by Pub Year and TD Type}

\begin{figure*}
  \centering
  \includegraphics[width=\textwidth]{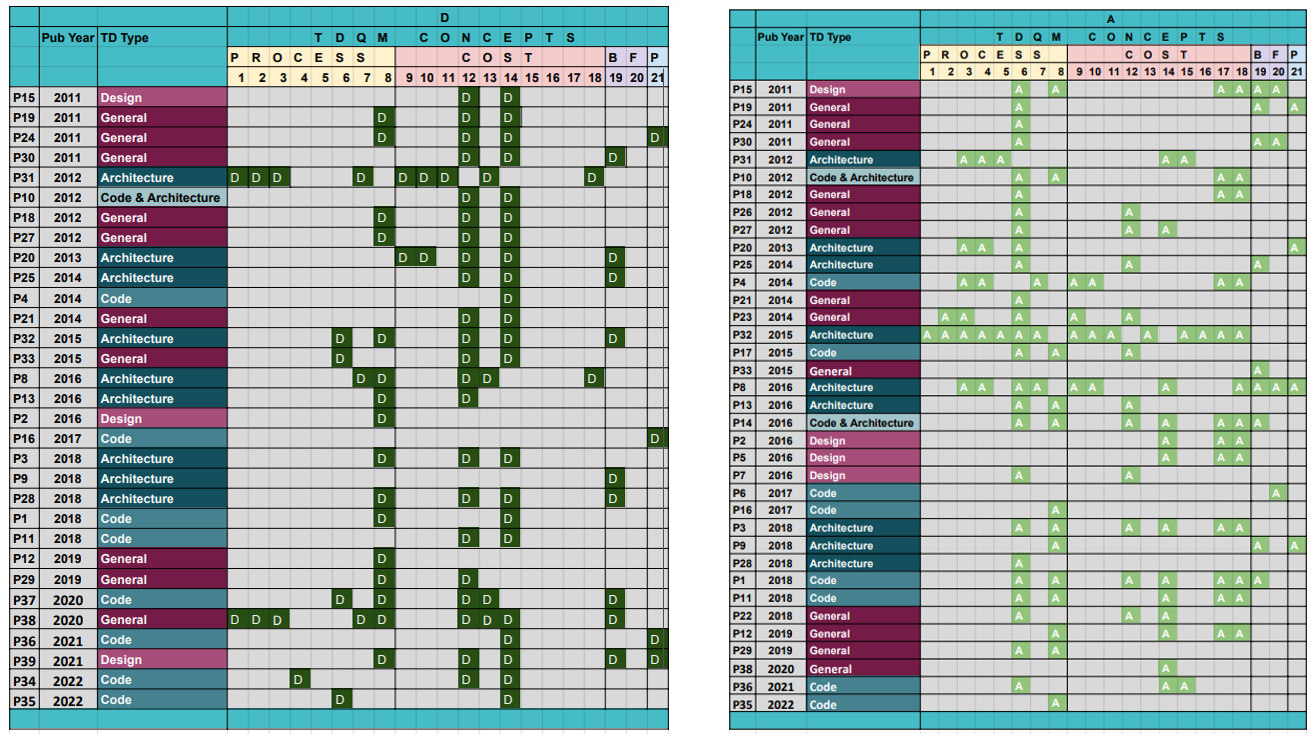}
  \caption{Left: TDQM Approach Comparison Matrix - Compact View - \emph{Direct (D)} Mappings sorted by Pub Year and TD Type | Right: TDQM Approach Comparison Matrix - Compact View - \emph{Associated (A)} Mappings sorted by Pub Year and TD Type}
  \label{fig:2DvsA}
\end{figure*}

In this slightly different representation of the TDQM Approach Comparison Matrix (See Right, Figure \ref{fig:2DvsA}), where the sorting of the approaches is based on first the Publication Year and then the TD Type, it is apparent that `refactoring' has been a popular discussion topic throughout. However, we see only \emph{Associated (A)} mappings with the `Refactoring Step.' `Interest Probability' has been a far less discussed topic than many of the TDQM Concepts. However, in comparison to the previous representation of the Matrix, we can see more mappings covering this concept in this representation. This means that there have been approaches that had some association with the concept even though they might not have explicitly discussed it, i.e., no \emph{Direct (D)} mapping. `Rework Cost not associated with TD' and 'New Code Cost not associated with TD' received equal attention. However, we see more mappings for `Rework Cost associated with TD' and 'New Code Cost associated with TD' compared to the previous representation of the Matrix. The `TD Item' concept has generally been popular, similar to the previous representation.  

\subsection{Combined Mappings (1)}

\subsubsection{Similarities in terms of what TDQM Concepts are being quantified}
\label{sec:similarities}


A combined mapping combines mappings \emph{D and A} into a single mapping. Here we utilize the \emph{Combined Mapping} to identify similarities and differences between quantification approaches. 
Nevertheless, if a detailed comparison is required, one could also compare the quantification approaches using the more detailed mappings \emph{D} and \emph{A}. 

\begin{figure*}
  \centering
  \includegraphics[width=\textwidth]{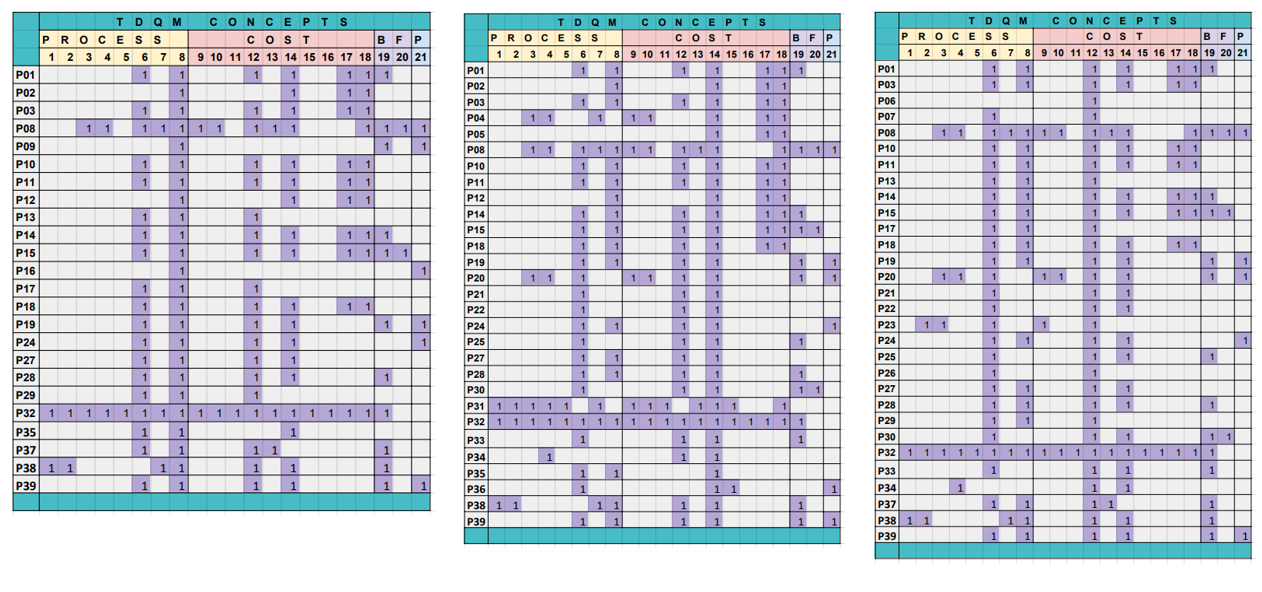}
  \caption{Similarities in terms of TDQM Concepts | Left: Approaches quantifying TD Item, Middle: Approaches quantifying Interest, Right: Approaches quantifying Refactoring Cost}
  \label{fig:similarities}
\end{figure*}

Figure \ref{fig:similarities} illustrates a few examples where we compared the 39 quantification approaches concerning an individual TDQM concept to identify \textbf{similarities in terms of what they quantify}. The examples are given for the TDQM Concepts, \emph{TD Item, Interest and Refactoring Cost}.

\subsubsection{Ordering quantification approaches based on coverage of TDQM Mappings}
\label{sec:ordering}

\begin{figure*}
  \centering
  \includegraphics[width=\textwidth]{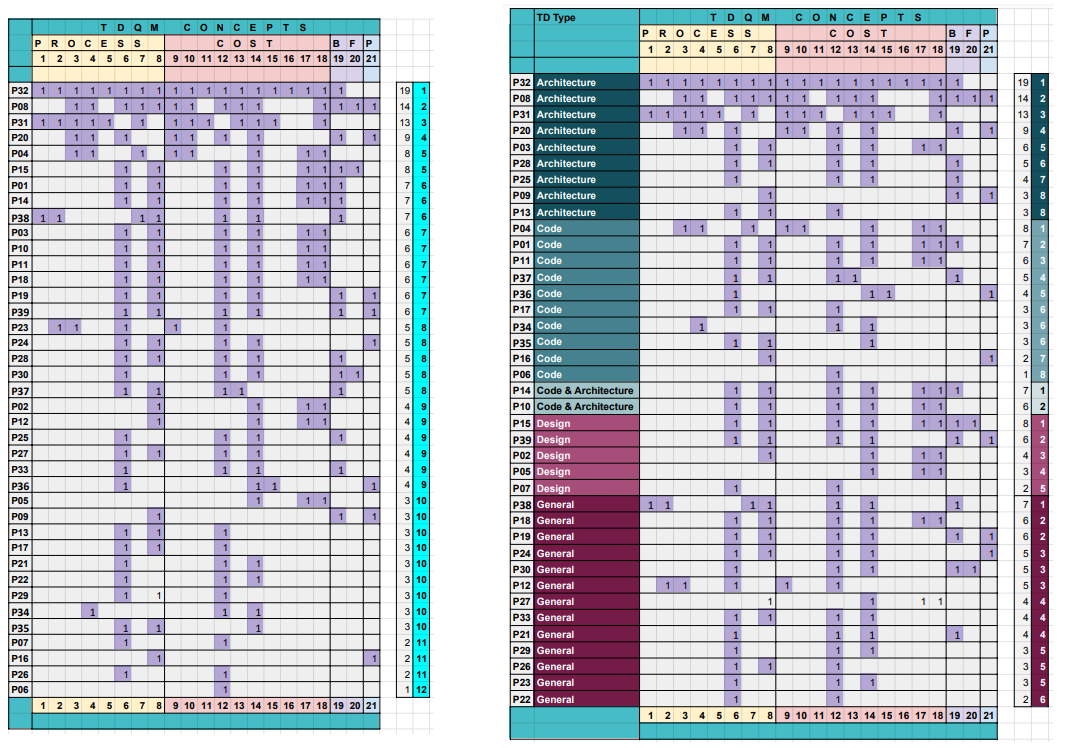}
  \caption{Left: Approaches ordered based on the mapping score for TDQM Concepts | Right: Approaches ordered within each TD type based on the mapping score for TDQM Concepts}
  \label{fig:ranking}
\end{figure*}

We attempted to order quantification approaches based on the coverage of TDQM concepts. (i.e., depending on how many TDQM concepts the particular quantification approach covered). We utilized the \emph{`Combined'} mapping for this purpose. The result can be seen in Figure \ref{fig:ranking}. 

See Left of Figure \ref{fig:ranking}, P32 covers 19 of the TDQM Concepts except 'Benefit of taking TD' and `Interest Probability' (See Figure \ref{fig:viz comparison} for the Legend for the TDQM concepts). Hence, it receives the topmost position in the ordering. P08 and P31 achieve positions 2 and 3, covering fourteen and thirteen TDQM concepts, respectively. 

See Right of Figure \ref{fig:ranking}, here the quantification approaches are ordered within each TD Type. P8 becomes the second in position within the TD type 'Architecture' following P32 in position 1. For Code TD, P04 achieves position 1 while P14, P15, and P38 receive the first position within TD types `Code and Architecture', `Design', and 'General', respectively.

However, if the ordering of quantification approaches in this manner is valuable and if it facilitates selecting an approach fit for the purpose is arguable. Although, what becomes evident is that TDQM allows exploring possibilities of ordering approaches based on the coverage of concepts. Thereby, TDQM forms the basis for assessing quantification approaches for their comprehensiveness. Ordering quantification approaches is further discussed in Section \ref{sec:discussion}.

\subsection{Trends}
\label{sec:trends}

\begin{figure}
  \centering
  \includegraphics[scale=0.5]{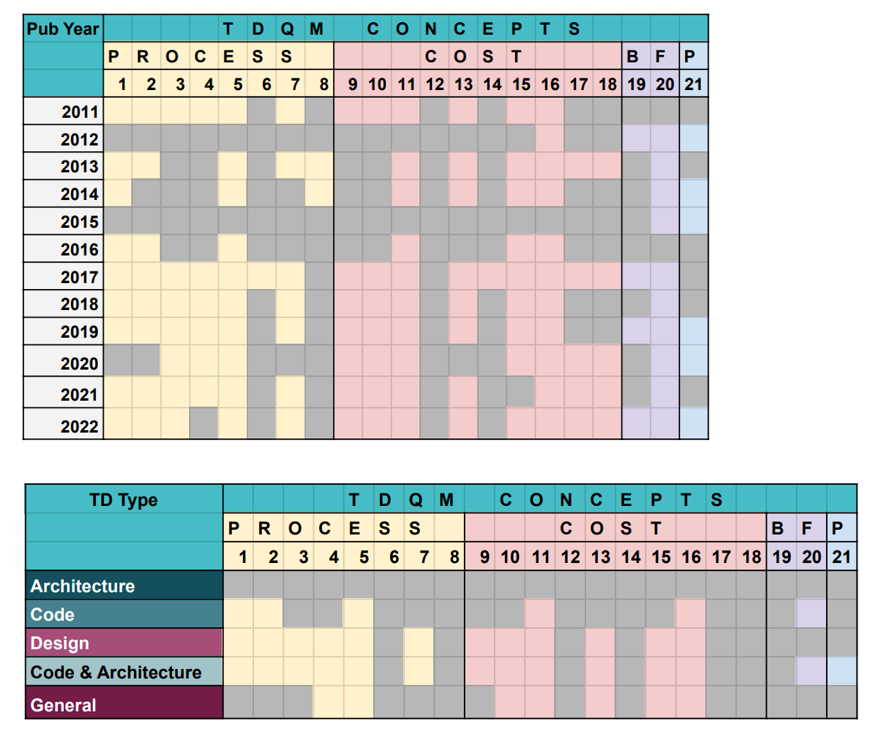}
  \caption{Trends | Top: Publication Year Vs TDQM Mappings, Bottom: TD Type Vs TDQM Mappings}
  \label{fig:trends}
\end{figure}

Figure \ref{fig:trends} shows trends that can be identified concerning the coverage of TDQM concepts (see Figure \ref{fig:viz comparison} for the Legend for TDQM concepts 1-21). Regarding the \emph{Publication year}, we can see that 2012, 2015, and 2016 seem to be the years where most of the TDQM Concepts have been discussed. In terms of the \emph{TD Type}, we can see that \emph{`Architectural TD'} covers all 21 TDQM concepts, while Code TD is the next TD Type that discusses 15 TDQM Concepts.

\subsection{Overall Results --- Number of approaches per TDQM mapping 1, D, A and C}

\begin{figure*}
  \centering
  \includegraphics[width=\textwidth]{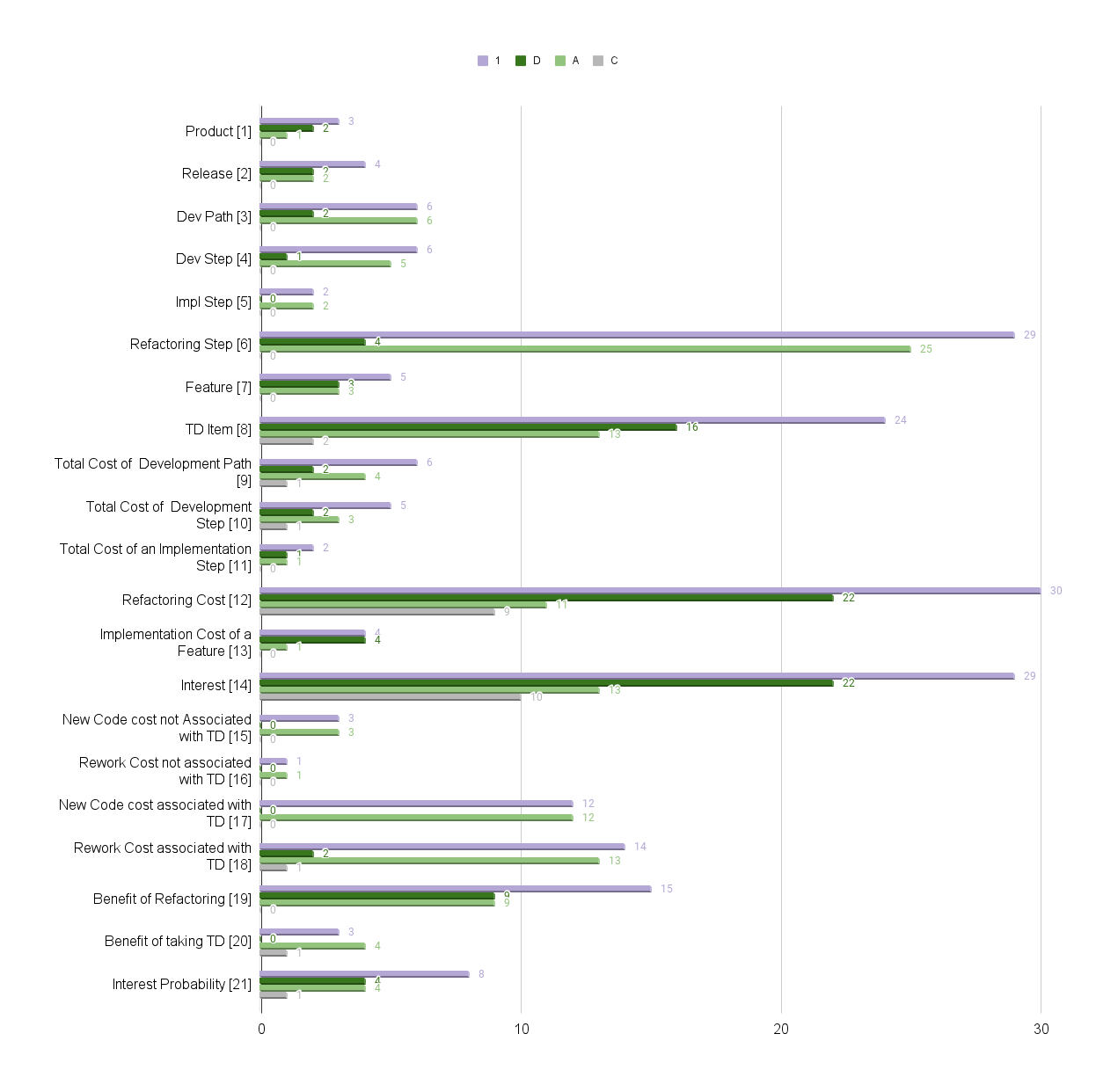}
  \caption{Number of approaches per TDQM mapping 1, D, A and C}
  \label{fig:num_app_per_mappings}
\end{figure*}

Figure \ref{fig:num_app_per_mappings} shows the overall result of the number of approaches per TDQM mappings \emph{1 - Combined, D - Direct , A - Associated} and \emph{C - Contributes} for the quantification approaches found in our mapping study complete dataset. 

\subsubsection{Direct Mappings (D)}
TDQM concepts \emph{Refactoring Cost, TD Interest, and TD Item} received the highest numbers of approaches that had Direct (D) mappings to TDQM concepts --- 22, 22, and 16, respectively. 

\subsubsection{Associated Mappings (A)}
TDQM concepts \emph{TD Interest, TD Item, and Rework cost associated with TD} received an equal number of approaches (13) that had Associated (A) mappings. New Code cost associated with TD and Refactoring Cost had 12 and 11 approaches with Associated mappings. 

\subsubsection{Contributes Mappings (C)}
The TDQM concepts \emph{TD Interest} and \emph{Refactoring Cost} received the highest mappings for the mapping type `Contributes. \emph{TD Interest} had nine approaches while \emph{Refactoring Cost} had ten approaches. 

\newpage
\section{Discussion}
\label{sec:discussion}

\subsection{TDQM Concepts}

Among the TDQM concepts discussed commonly in the different quantification approaches, \emph{TD Item, Cost of Refactoring and Interest} received the highest numbers of mappings for the primary studies in our dataset (16, 22, and 22 Direct mappings and 13, 11 and 13 Associated mappings respectively). We discuss below some findings concerning these three individual TDQM concepts.

\subsubsection{TD Items are Development Artefacts}

The 16162 model \cite{avgeriou2016managing} discussed in Section \ref{sec:related_work} associates the concept of a TD Item with one or more artifacts of the software development process as code, test, or documentation. Using TDQM and the mappings from the quantification approaches to TDQM, we further validate the association made in the 16162 model through our findings. 

\begin{figure*}
  \centering
  \includegraphics[scale=0.75]{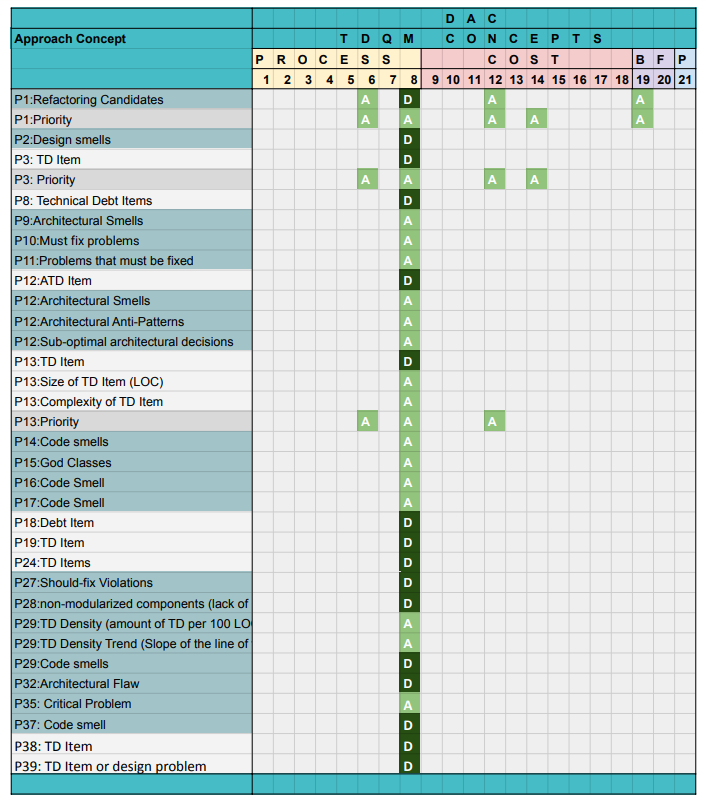}
  \caption{Approach Concepts that refer to Development Artefacts that map to the TDQM Concept `TD Item'}
  \label{fig:TD_Item_artifact}
\end{figure*}

Figure \ref{fig:TD_Item_artifact} shows the mappings \emph{D} and \emph{A} for `TD Item.' Most of the mappings referred to a development artefact in the quantification approaches. For example, P1 - `Refactoring candidates', P2 - `Design Smells', P9, P12 - `Architectural Smells' and `Architectural Anti-Patterns', P14, P16, P17, P29 and P37 - Code Smells, P15 - `God Classes'. 
However, some approaches used the term 'TD Item' instead of specifying a development artefact. Examples include P3, P8, P12, P13, P18, P19, P24, P38, and P39. 

Although not discussed in the 16162 model, concepts such as `priority' were identified as associated with the TDQM concept `TD Item' since each TD Item could have a priority that indicates that they need to be prioritized to be eliminated via a \emph{Refactoring Step} (e.g., P1, P3, P13). 

\subsubsection{Refactoring Cost is the same as `TD Principal'}

\begin{figure*}
  \centering
  \includegraphics[scale=0.5]{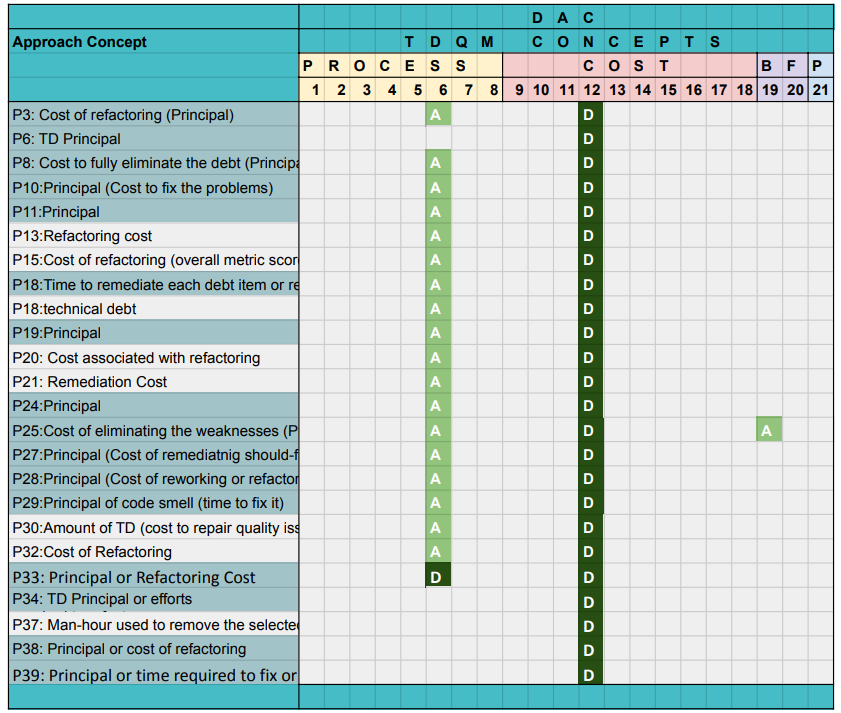}
  \caption{Approach Concepts that refer to `Principal' that map to the TDQM Concept `Refactoring Cost'}
  \label{fig:Refact_cost_vs_Principal}
\end{figure*}

An observation made via the TDQM mappings is that quantification approaches that refer to `TD Principal' refer to the same concept, `Refactoring Cost.' (See Figure \ref{fig:Refact_cost_vs_Principal}). Examples include P3, P6, P8, P10, P11, and P18. Further describing one of the examples, P27 refers to Principal as the `cost of remediating should-fix violations,' which is mapped to the TDQM concept `Refactoring Cost.' P33 and P38 explicitly mention `Principal' or `Cost of refactoring.'

Most approaches commonly use the concept `Principal' with the phenomenon 'Technical Debt (TD)' since it refers to the sum of money lent or invested on which the interest is paid, according to the financial definition. However, if the Principal refers to `Refactoring Cost,' is arguable as it does not refer to the amount borrowed originally since the cost of refactoring can become higher than the amount of time or effort borrowed in the first place by taking TD.

\subsubsection{Interest Decomposition for code-related TD Types}

\begin{figure*}
  \centering
  \includegraphics[width=\textwidth]{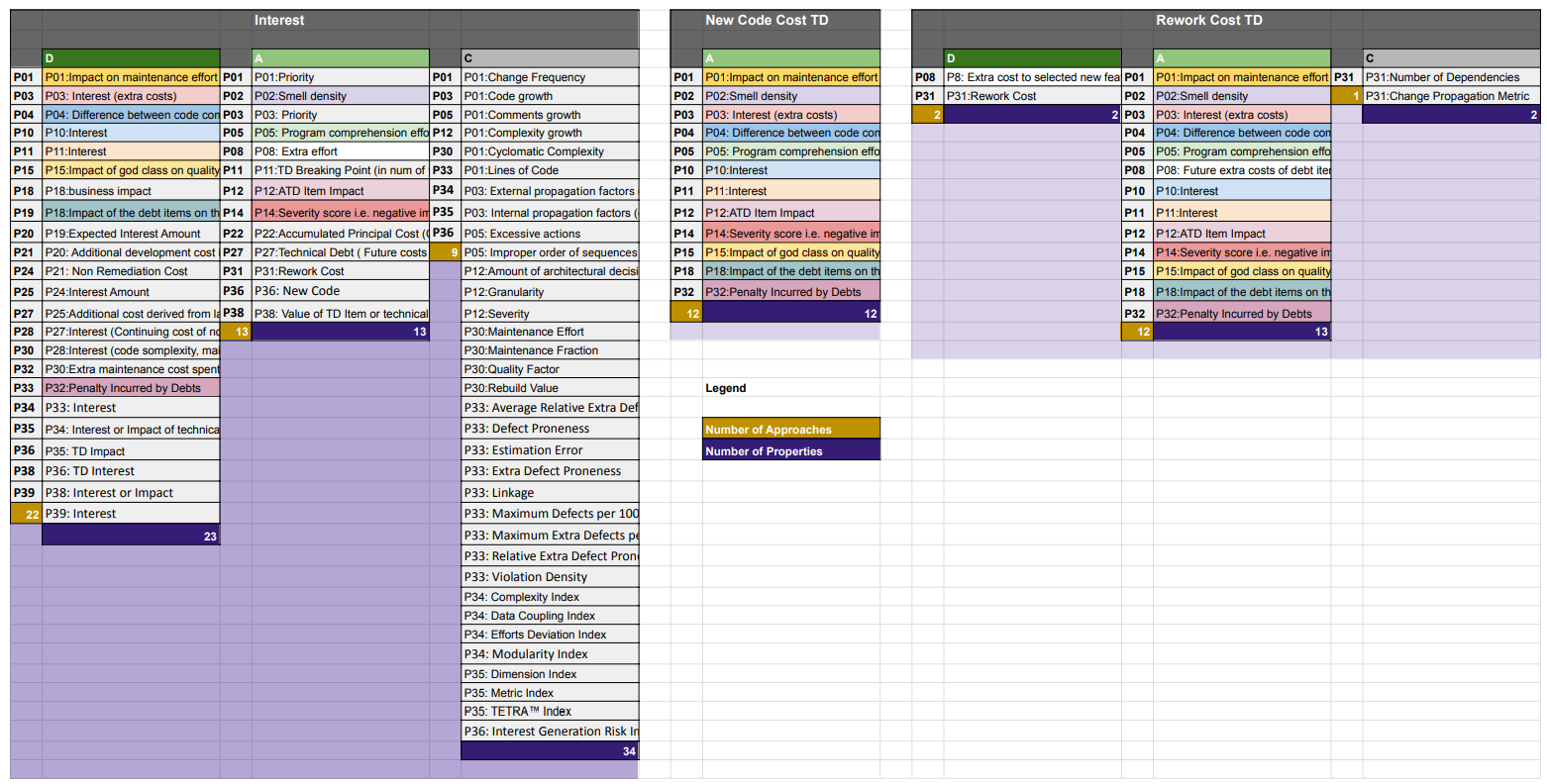}
  \caption{\emph{D}, \emph{A} and \emph{C} mappings for TD Interest Decomposition: New Code Cost and Rework Cost}
  \label{fig:int_decomp}
\end{figure*}

The decomposition for TD Interest described in TDQM (for code-related types of TD), received multiple mappings (See Figure \ref{fig:int_decomp}). Some mappings resulted from the approaches having the concept explicitly discussed, while some mappings were from the derivation of the association through our understanding. 

`Rework cost associated with TD' was explicitly discussed (i.e., had mapping D) in 2 quantification approaches, while 13 approaches had some association with the concept (i.e., mapping A). `New Code cost associated with TD' did not have Direct (D) mappings. Although, 12 approaches had some association (i.e., mapping A) with the concept. 

Through the color coding in Figure \ref{fig:int_decomp}, we illustrate the relationship, how the properties found for either a \emph{Direct or Associated} mapping with TDQM for TD Interest is further decomposed into either \emph{`New code cost or Rework cost associated with TD}. One example is `P32: Penalty Incurred by Debts', which we also discussed previously in Section \ref{example application of TDQM}. Another example is, 'P15: Impact of god class on quality attributes (defect likelihood, change likelihood)', which indicates that it could involve either new code or rework. Hence, it is worth exploring further the decomposition for TD Interest, i.e., the relationship between New Code, Rework and TD. 

\subsection{Contributions of TDQM}

TDQM contributes to existing research in multiple ways. We highlight the main contributions and discuss them in the following subsections.

\subsubsection{Systematically comparing and evaluating quantification approaches within the TDQM Approach Comparison Matrix}

Section \ref{example application of TDQM} illustrated how TDQM can be utilized to compare and evaluate two quantification approaches. Similarly, the \emph{TDQM Approach Comparison Matrix} can be utilized to compare and evaluate more than two approaches at the same time, increasing the efficiency in the comparisons which is evident through the study results in Section \ref{sec:results:tdqm_app}. Quantification approaches could be compared either in the \emph{Detailed View} or in the \emph{Compact View} of the Matrix depending on the level of detail one would like examine. Section \ref{sec:similarities} illustrates comparing approaches for similarities based on the \emph{Combined (1)} mapping.  

\subsubsection{Classifying and visualizing trends concerning quantification approaches}

The \emph{TDQM Approach Comparison Matrix} could be utilized to visualize different classifications of quantification approaches; one being the visualization of trends. Section \ref{sec:trends} discusses results identifying such trends. Apart from the visualization of trends multiple other classifications and visualizations of the Matrix can be found in Section \ref{sec:results:tdqm_app} and in the Replication Package\footnotemark[2]. 

\subsubsection{Assessing quantification approaches through ordering them based on the coverage of TDQM Concepts}
\label{sec:ordering_disc}

TDQM can be utilized to order quantification approaches based on the coverage of TDQM concepts. The coverage of TDQM concepts implies how much a quantification approach comprehends TD quantification. Hence, this forms the basis for assessing quantification approaches for their comprehensiveness. In Figure \ref{fig:ranking}, we illustrated two ways of ordering quantification approaches found in our mapping study based on the coverage of TDQM concepts. 

As illustrated in the Left side of Figure \ref{fig:ranking}, one way of ordering quantification approaches is ordering all the approaches based on the coverage of concepts regardless of the TD type. The approach at the top of Figure \ref{fig:ranking}, P32, covers the highest number of TDQM concepts. Therefore, it can be regarded as the most comprehensive approach. 

The other way is first categorizing the approaches based on the TD type and then ordering them within the TD type for their coverage of TDQM concepts as illustrated in the Right side of Figure \ref{fig:ranking}. The approach that covers the highest number of TDQM concepts for each TD type can be seen on the top of each block for the TD Type. One could argue that this is the comparatively better way of ordering approaches because approaches from the same type of TD can be more similar in quantifying a given individual TDQM concept. 

Suppose we want to determine the usefulness of the quantification approaches in TDM decision-making. We can still consider the ordering of approaches based on the coverage of TDQM concepts to be helpful in this evaluation since TDQM concepts provide an idea of what TDM decisions could be made given the quantification of the concepts. For example, an approach that covers `Refactoring Cost' and 'Benefit of Refactoring' helps make TDM decisions, such as \emph{when to refactor}. 

Suppose the approaches are evaluated for their feasibility in practice, for example, in a specific development environment. In that case, we cannot determine how practical these approaches are, based on the coverage of the TDQM mappings.

\subsubsection{Serving as a reference point to develop new quantification approaches}

Existing approaches have no consensus among them regarding what they quantify and what TDM decisions they support. Hence, it is clear that existing approaches cannot serve as a reference point to understand and build new TD quantification approaches. TDQM fills this gap by consolidating the important concepts related to TD Quantification and also improving the comprehension of TD quantification by showing the relationships between these concepts via the TDQM conceptual model (See Figure \ref{fig:Concept_Map_model}).
 
\subsection{Limitations of TDQM}

The initial mapping study was conducted for all types of TD, e.g., Code, Design, Architectural, Requirements, Documentation, Test, and Process. However, during the development of TDQM, we chose to develop the model for the quantification of TD types related to software code such as, Code, Design, Architectural, and General (inferred as code-related).' Trying to develop a model for all TD types at once might not have been feasible. Therefore, we had to pick a starting point. Since most static analysis tools used in the \emph{identification of TD} focus on code-related TD types, it motivated us to start modeling the quantification of TD for code-related types of TD. Therefore, even though our mapping study resulted in 129 primary studies overall, we applied TDQM to only the 39 studies which pertained to the code-related TD types (See Table \ref{tab:typesoftd_SMS_it}).

\subsection{Threats to Validity of the Systematic Mapping Study}

\subsubsection{Identification of Primary Studies}

Threats to identifying primary studies, i.e., missing articles, could apply to our mapping study during its search phase. The search string was developed in multiple iterations before we finalized a satisfactory search string and then tested for its accuracy. We piloted the search string multiple times with one of the major databases, SCOPUS, and then checked our results with the reference set of articles we retrieved from the TD conference proceedings in 2018 ad 2019. All the articles from our initial manual search could be found in SCOPUS. Therefore, we could easily verify our search string with the reference set of articles. We used keywords and their synonyms and wildcards (*) to capture possible variations of the keywords, for example, plurals and verb conjugations. We applied the search query to the title, abstract, and keywords to increase the probability of finding all relevant articles. 

Furthermore, in the first iteration of the mapping study, we did not limit our search to a particular period, although the retrieved articles spanned between 2011-2020. To eliminate the threat of missing articles, we conducted reference snowballing on the final set of articles from the screening. We followed this step so that any articles not captured by the search string would be found during this step.

The second iteration of the mapping study was conducted from 2020-2022 to keep it up to date. The same process was followed; for example, the search string and the inclusion and exclusion criteria developed in the initial mapping study were also used here. 

\subsection{Threats to Validity of the Development of TDQM}

\subsubsection{Researcher Bias}

Three researchers were involved in the creation of our model. The first author investigated a subset of quantification approaches derived from the first iteration of the mapping study in order to build the model in the first place. The model was then developed in iterations while having ongoing discussions with the other two researchers. Once the model was developed with enough confidence, it was applied to another subset of quantification approaches to evaluate its feasibility. A fourth researcher reviewed our model, providing feedback from the viewpoint of a person who was not involved in creating the model. We also asked for feedback from our research group. We reduced subjectivity that might have been introduced in the process by following those verification steps.

\subsubsection{Construct Validity}

Construct validity applies to TDQM in multiple ways. These include;

\begin{itemize}
    \item If another set of researchers will develop a similar or different model given the set of Primary Studies identified via our first iteration of the mapping study 
    \item If another set of researchers do the same mapping for a given approach i.e., assigning D, A, C
    \item If a concept related to TD quantification is a valid concept to be included in TDQM if only one quantification approach discusses it
    \item If there could be concepts missing in the model that might still be important for modeling the quantification of TD
\end{itemize}

Although such possibilities might exist, we are confident with how we mitigated them. Our model was informed by previous literature, i.e., past models of TD and our experience as software developers in the industry. Most of the important parts of the model were captured by many quantification approaches we applied TDQM to (e.g., Cost of Refactoring, TD Interest, and TD Item captured by 21, 21, and 16 approaches, respectively); this theoretically validated the model. Additionally, we have gained more confidence in our model as we applied the model to multiple quantification approaches in two rounds of the mapping study. --- The initial round (Iteration 1) which included primary studies from 2011-2020, and the next round (Iteration 2) where we updated the mapping study with new results for 2020-2022. The model was applied to all the unique TD quantification approaches found in both iterations of the mapping study; before and after the development of TDQM. The model also did not change in the second iteration of the study. Therefore, this gives confidence that the model sufficiently captures what is required for modeling the quantification of TD for code-related types of TD; the selection of concepts for the model and the relationships between them.

\subsubsection{External Validity}

The motivation to develop the model came from identifying the need to efficiently compare and evaluate quantification approaches found in our mapping study. Thus, the model was developed with knowledge of existing quantification approaches. Once the model was developed, we applied it to all unique quantification approaches found in the mapping study, categorized into Code, Design, Architectural, and General types of TD. The model was successfully applied to all 33 code-related TD quantification approaches from the first iteration of the mapping study. Additionally, we applied the model to unique quantification approaches resulting from our mapping study's extension, which included 6 more approaches from Code, Design, Architectural and General types of TD to the dataset. TDQM was successfully applied to all of these new quantification approaches as well, increasing the total number of approaches that TDQM was applicable to 39. Most importantly, the model did not change during the second iteration of the mapping study. Therefore, we believe that the model generally applies to TD quantification approaches for code-related types of TD. We assume that our mapping study captured all studies that propose a quantification approach, although we acknowledge that this might not be the case.

\subsection{Threats to Validity of the Application of TDQM}

\subsubsection{Researcher Bias}
The first author applied TDQM to the quantification approaches found via the mapping study. Therefore, we acknowledge that a researcher bias might be introduced by the overall knowledge the first author already gained, being involved as the main researcher for all three research questions presented in this paper. However, the application of the model and the results were discussed and verified among the other authors and presented in research groups and seminars
for feedback. Additionally, our work \cite{10.1145/3563768.3565553} was presented at the SPLASH Doctoral Symposium 2022. Therefore, we believe that this threat has been sufficiently mitigated.

\subsection{Future Work and Implications to Researchers and Practitioners}

TDQM was developed to model the quantification of code-related types of TD such as Code, Design, and Architecture TD. Therefore, our future work will explore how TDQM can be extended to other types of TD. However, we chose Requirements TD as a non-code-related type of TD since Requirements can be directly connected with the existing TDQM concept `Feature' (i. e., requirements are developed into features). 

Our working definition for Requirements Debt is; \emph{"the impact or interest (e.g., in terms of costs) of sub-optimally implementing (not implementing or partially or incorrectly implementing) requirements"}. We are aware that Requirements Debt can be introduced in the process of eliciting requirements as well as when developing features from a System Requirements Specification (SRS).  
We plan to investigate this with a case study. The research question we would like to answer via the case study is; \emph{"How can we quantify TD Interest for non-code related types of TD such as Requirements Debt?"}. 

However, we invite researchers to further investigate the quantification of non-code-related types of TD such as Test, Documentation, and Process TD.

Another area that we plan to conduct our future work is by experimenting with parts of our model. We plan to investigate the relationship between TD Items, New Code and Rework as we think that New Code and Rework could possibly be used as a proxy to quantify TD Interest as they are constituents of TD Interest (for code-related types of TD), according to our model. 

As part of our future work, we have developed a calculation model and a decision model, which we plan to include in our future publications. Nevertheless, we invite researchers and practitioners to investigate further how TDQM can support TDM decision-making.

\section{Conclusion}
\label{sec:conclusion}

We conducted a systematic mapping study to determine what approaches to TD quantification have been proposed in the research literature. Although the mapping study resulted in multiple approaches to TD quantification, it was unclear if these quantification approaches quantified the same thing and if they supported similar decisions concerning TDM. This indicated that there was no consensus among the approaches and therefore, it is difficult to sensibly and efficiently compare and evaluate the various quantification approaches. 

To solve this problem, we developed TDQM, a model that captures the important concepts related to TD quantification and illustrates their relationships. TDQM allows representing quantification approaches via a common uniform representation --- the \emph{TDQM Approach Comparison Matrix} that facilitates systematic and efficient comparisons and evaluations among quantification approaches. Additionally, TDQM serves as a reference point to comprehend TD quantification and to develop new quantification approaches based on it.

We demonstrate the use and value of TDQM by applying it to compare and evaluate two example quantification approaches; Nord et al. and Kazman et. al.. We then evaluate TDQM by applying it to classify, compare and evaluate 39 quantification approaches found during our mapping study --- 33 prior to and 6 after the development of TDQM. This study's results indicate that TDQM applies to quantification approaches that appear quite different in form, serving as a platform to efficiently compare and evaluate them. 

With the use of the \emph{TDQM Approach Comparison Matrix}, we were also able to derive useful results for the evaluation of the quantification approaches found in our mapping study. For example, we found that among the TDQM concepts discussed in the different quantification approaches, \emph{TD Item, Cost of Refactoring and Interest} received the highest numbers of mappings; 16, 22, and 22 Direct mappings and 13, 11 and 13 Associated mappings respectively. Furthermore, we identified trends among the quantification approaches. For example, Architecture Debt was the type of TD that covered most of the concepts of TDQM, and the publication years 2012, 2015, and 2016 were where TD quantification (i.e., TDQM concepts) was mainly discussed. 

The use of TDQM also highlighted aspects of TD quantification that have not been extensively explored previously, for example, `New Code Cost' and `Rework Cost.' In our future work, we plan to explore the relationship between TD Items and the concepts of `New Code Cost' and `Rework Cost,' which we identified as constituents of TD Interest, during modeling the quantification of TD for code-related types of TD. Furthermore, we plan to explore the possibility of extending TDQM to non-code-related types of TD such as Requirements Debt.


\bibliographystyle{ACM-Reference-Format}
\bibliography{References}







\end{document}